\documentclass[preprint2]{aastex} % demo
\usepackage{graphicx}
\usepackage{amsmath}
\usepackage{natbib}
\usepackage[colorlinks=true,citecolor=blue,breaklinks=true,linktocpage=true]{hyperref}
\bibpunct{(}{)}{;}{a}{}{,} %% natbib format like A&A and ApJ
\usepackage[switch,pagewise]{lineno}
\usepackage{xspace}
\usepackage{hyperref}
\usepackage{xcolor} 
\newcommand{\alfven}{Alfv\'en\xspace}
\newcommand{\unit}[1]{\ensuremath{\,\mathrm {#1}}}  % text in math mode
\renewcommand{\d}{\ensuremath{\text{d}}}  % differential in formulate
\newcommand{\A}{\ensuremath{\text{A}}}  %
\newcommand{\s}{\ensuremath{\text{s}}} 
\newcommand{\T}{\ensuremath{\text{T}}}  
 
\renewcommand{\i}{\ensuremath{\text{i}}}  
\newcommand{\e}{\ensuremath{\text{e}}} 
\newcommand{\figref}[1]{Figure~\ref{#1}}
\newcommand{\secref}[1]{Section~\ref{#1}}
\renewcommand{\eqref}[1]{Equation~\ref{#1}}
\newcommand{\bvec}[1]{\ensuremath{\boldsymbol{\mathbf{#1}}}}  
\renewcommand{\div}[1]{\nabla\cdot #1} % for divergence
\newcommand{\curl}[1]{\nabla\times #1} % for curl
\newcommand{\grad}[1]{\nabla#1} % for gradient
\newcommand{\singlequote}[1]{\lq{#1}\rq}
\renewcommand{\deg}{\ensuremath{^\circ}}
\newcommand{\RN}[1]{\uppercase\expandafter{\romannumeral#1}}
\newcommand{\bigO}[1]{\ensuremath{\mathcal{O}{(#1)}}}
\newcommand{\RR}{\ensuremath{\mathcal{R}}}
\newcommand{\feix}{\ion{Fe}{9} $\lambda171.073$ \AA{}\xspace}

\newcommand{\kri}{\ensuremath{|\kappa_{r\i}|}\xspace}
\newcommand{\kre}{\ensuremath{\kappa_{r\e}}\xspace}
\shorttitle{standing kink wave}
\shortauthors{Yuan et al.}

\begin{document}

\title{Forward Modelling of Standing Kink Modes in Coronal Loops \RN{1}. Synthetic Views}
\author{Ding Yuan\altaffilmark{1,2,3}}
\email{DYuan2@uclan.ac.uk}
\author{Tom Van Doorsselaere\altaffilmark{1}}
\altaffiltext{1}{Centre for mathematical Plasma Astrophysics, Department of Mathematics, KU Leuven, Celestijnenlaan 200B bus 2400, B-3001 Leuven, Belgium}
\altaffiltext{2}{Jeremiah Horrocks Institute, University of Central Lancashire, Preston PR1 2HE, UK}
\altaffiltext{3}{Key Laboratory of Solar Activity, National Astronomical Observatories, Chinese Academy of Sciences, Beijing, 100012, China}
\begin{abstract}
Kink magnetohydrodynamic (MHD) waves are frequently observed in various magnetic structures of the solar atmosphere. They may contribute significantly to coronal heating and could be used as a tool to diagnose the solar plasma. In this study, we synthesise the \feix emission of a coronal loop supporting a standing kink MHD mode. The kink MHD wave solution of a plasma cylinder is mapped into a semi-torus structure to simulate a curved coronal loop. We decompose the solution into a quasi-rigid kink motion and a quadrupole term, which dominate the plasma inside and outside the flux tube, respectively. At the loop edges, the line-of-sight integrates relatively more ambient plasma, and the background emission becomes significant. The plasma motion associated with the quadrupole term causes spectral line broadening and emission suppression. The periodic intensity suppression will modulate the integrated intensity and the effective loop width, which both exhibit oscillatory variations at half of the kink period. The quadrupole term can be directly observed as a pendular motion at front view.
\end{abstract}


%\begin{comment}
\section{Introduction}
\label{sec:intro}

In the past decade, significant progress has been achieved in probing the heating and seismological roles of magnetohydrodynamic (MHD) waves in the solar atmosphere \citep[see reviews by][]{nakariakov2005,liu2014,jess2015,arregui2015}. Among the MHD wave modes, the slow magnetoacoustic mode propagates anisotropically in a low $\beta$ uniform plasma; the wave energy flows predominantly along the magnetic field line, e.g., \citet{demoortel2002b,demoortel2002a, wang2009a,wang2009b, yuan2012sm,kumar2013,kumar2015,fang2015}. While fast magnetoacoustic mode could propagate to any direction relative to the magnetic field, i.e., either parallel, perpendicular or oblique, therefore, they are commonly waveguided in a variety of magnetic structures through reflections and refractions. They may couple with \alfven wave and exhibit mixed wave properties in forms of standing transverse oscillations of coronal loops \citep{aschwanden2011,nistico2013,verwichte2013}; large-scale coronal propagating fronts across the whole solar disk \citep{ofman2002, liu2010,guo2015}; quasi-periodic fast wave trains along magnetic funnels \citep{liu2012,yuan2013fw,pascoe2013,nistico2014}; fast wave pulses across randomly structured plasma \citep{yuan2015rs}; ubiquitous propagating kink waves in the entire corona \citep{tomczyk2007} and coronal holes \citep{thurgood2014,morton2015}.

Kink waves \citep[the $m=1$ mode,][]{edwin1982,edwin1983,ruderman2003,erdelyi2009,goossens2014} were initially observed in active region loops in the Extreme Ultraviolet (EUV) channels of the Transition Region and Coronal Explorer \citep[TRACE,][]{nakariakov1999,aschwanden1999}. The coronal loops were observed to oscillate transversely with amplitudes at megametre scale in response to flares, i.e., the associated mass ejections \citep{schrijver2002,zimovets2015}, filament destabilizations \citep{schrijver2002}, magnetic reconnection \citep{he2009}, or vortex shedding \citep{nakariakov2009}. Recently, \citet{nistico2013} and \citet{anfinogentov2013} detected low-amplitude (sub-megametre scale) kink oscillations of coronal loops. Kink waves in this category last for dozens of wave cycles without significant damping, and are apparently not associated with any explosive events \citep{anfinogentov2013}. Transverse oscillatory motions were also observed in chromospheric spicules \citep{okamoto2011,morton2014}, chromospheric mottles \citep{kuridze2012}, filament threads \citep{lin2007,lin2009}, large prominences \citep{tripathi2009,hershaw2011,arregui2012}, polar plumes \citep{thurgood2014}, coronal rain \citep{antolin2011}, helmet streamers \citep{chen2010,chen2011}, and even coronal mass ejections \citep{lee2015}.

Fundamental (global) standing kink modes are frequently observed in closed coronal loops \citep{vandoorsselaere2009,ruderman2009}. The period of coronal transverse waves ranges from 2 min to 33 min; and the damping time has a similar time scale \citep{aschwanden2002,white2012}. The curved coronal loops are normally assumed to be approximately co-planar; the loop plane intrinsically defines horizontally and vertically polarized kink waves about the loop axis \citep{ruderman2009b}. Horizontal kink waves are more frequently observed, e.g., \citet{nakariakov1999,schrijver2002,aschwanden2002,zimovets2015}; while vertical kink waves were only reported in a limited number of cases, e.g., \citet{wang2004,verwichte2006,selwa2007,selwa2010,selwa2011,white2012b,kim2014}.

The main interest in standing kink modes of coronal loops arises mainly from their role in diagnosing the coronal plasma via MHD seismology \citep{nakariakov2005,demoortel2012}. The standing kink mode could be used as a tool to infer magnetic field strength along a coronal loop \citep{nakariakov2001,verwichte2009,verwichte2010,verwichte2013}. \citet{verwichte2013b} measured the range of the density contrast and inhomogeneity layer thickness of coronal loops based on the period-damping time scaling law. \citet{demoortel2009} are the first to validate MHD seismology with three-dimensional numerical simulations, and showed that the inverted magnetic field strength agrees with the input magnetic field within a factor of about two. \citet{aschwanden2011} and \citet{verwichte2013} compared the seismological field and the \alfven-transit-time-averaged value in the potential field model, and found consistency within an order of magnitude. \citet{chen2015} performed MHD simulations using a realistic coronal model, and found that the excited coronal loop oscillations would be effectively used to infer the average magnetic field.

Kink MHD waves are highly-incompressible in the long wavelength limit and exhibit only quasi-rigid motions \citep{goossens2012}. Indeed, in a coronal loop the density (or temperature) perturbation by a kink mode is at the order of $10^{-3}$ or less of the equilibrium value. The observed intensity variations of coronal loops \citep[e.g.,][]{oshea2007,verwichte2009,verwichte2010} are ascribed to the column depth modulation introduced by the kink motion. \citet{cooper2003a,cooper2003b} performed line-of-sight (LOS) integration through the coronal loop plasma perturbed by MHD waves and demonstrated that intensity modulation could become significant in case of a kink mode, even though the plasma fluid compression is negligible.

Recently, \citet{goossens2014} showed that the kink mode solution could be decomposed into a quasi-rigid transverse motion and a rotational motion, which is detectable as Doppler velocity oscillations in optically thick lines. It confronts interpreting rotational motion as a signature of \alfven wave \citep{depontieu2012}. Therefore, forward modelling would significantly advance the knowledge of kink modes and resolve the dispute on whether a wave with observed rotational motion is a kink or \alfven wave \citep[e.g.,][]{vandoorsselaere2008}. Moreover, MHD seismology and wave energy estimation strongly rely on correct identification of the wave mode and accurate measurements of wave properties \citep{goossens2012,vandoorsselaere2014}.

Forward modelling is a novel approach that synthesizes the plasma emission observables \citep{antolin2013,yuan2015fm}. It basically converts analytical or numerical models into observables. Therefore, the inversion process (e.g., MHD seismology, MHD spectroscopy, helioseismology, x-ray tomography), which is originally ill-posed owing to the lack of sufficient constraints (or observables), could be better understood in the sense that knowledge of plasma properties is given a priori. \citet{gruszecki2012} studied the geometric integration of the plasma density of a fast sausage mode of a  plasma cylinder. \citet{demoortel2008} demonstrated that the damping rate measured in EUV emission intensity oscillations may not reflect the real damping of MHD waves. \citet{antolin2013} and \citet{antolin2014} considered the inhomogenuous plasma emission introduced by fast sausage modes and found that the LOS effect and spatial resolution would significantly modify the associated EUV emissions of coronal loops. \citet{yuan2015fm} found that the contribution function of atomic emission \citep{dere1997} could cause emission asymmetry for positive and negative temperature perturbations, and could even lead to the detection of half periodicity. 

In this study, we present the forward modelling study of standing kink modes of coronal loops. \secref{sec:model} gives the analytical solution of kink mode in a coronal loop and the numerical discretization for the forward modelling code\footnote{The FoMo code is available at \url{https://github.com/TomVeeDee/FoMo}}. \secref{sec:result} and \secref{sec:conclusion} present the results and conclusion, respectively.

\section{Model}
\label{sec:model}

\subsection{Standing kink mode}

In this paper, we study the standing kink wave in a plasma cylinder embedded in a uniform plasma. The magnetic field is parallel to the axis of the plasma cylinder (i.e., $z$-axis), $\bvec{B_0}=B_0 \bvec{\hat{z}}$. The equilibrium magnetic field $B_0$, plasma density $\rho_0$ and temperature $T_0$ are piecewise functions of the $r$-coordinate:
\begin{equation}
 B_0,\rho_0,T_0=\left\{
     \begin{array}{lr}
        B_\i,\rho_\i,T_\i,\text{ for $r \leq a$} \\
        B_\e,\rho_\e,T_\e,\text{ for $r>a$},
     \end{array}
   \right.
\end{equation}
where $a$ is the radius of the loop. Hereafter, we use subscript \singlequote{$\i$} and \singlequote{$\e$} to differentiate the internal and external equilibrium values of the loop system. 

The linearised ideal MHD equations give the perturbed quantities that deviate from the magnetostatic equilibrium \citep[see, e.g.,][]{ruderman2009}: 
\begin{align}
 \rho_1 &=-\div(\rho_0\bvec{\xi}), \label{eq:cont} \\
 \rho_0\frac{\partial^2\bvec{\xi}}{\partial t^2}&=-\grad{P_{\T1}} + \frac{1}{\mu_0}[ (\bvec{B_0}\cdot\grad)\bvec{b_1}+(\bvec{b_1}\cdot\grad)\bvec{B_0}],\label{eq:moment}\\ 
 \bvec{b_1}&=\curl(\bvec{\xi}\times\bvec{B_0}), \label{eq:mag}\\
 p_1-C_\s^2\rho_1&=\bvec{\xi}\cdot(C_\s^2\grad{\rho_0}-\grad{p_0})  \label{eq:entropy},
\end{align}
where $\bvec{\xi}$ is the Lagrangian displacement vector, $\rho_0$, $p_0$ and $\bvec{B_0}$ are the plasma density, pressure and magnetic field in equilibrium,  $\rho_1$, $p_1$ and $\bvec{b_1}$ are the perturbed plasma density, pressure and magnetic field, $P_{\T1}=p_1+\bvec{b_1}\cdot\bvec{B_0}/\mu_0$ is the perturbed total pressure, $\mu_0$ is the magnetic permeability in free space. A few typical speeds are defined to describe the loop system: $C_\s=\sqrt{\gamma p_0/\rho_0}$, $C_\A=B_0/\sqrt{\mu_0\rho_0}$, $C_\T=C_\A C_\s/\sqrt{C_\A^2+C_\s^2}$ are the acoustic, \alfven, and tube speed, respectively \citep{edwin1983}; and  $\omega_\s=C_\s k$, $\omega_\A=C_\A k$, $\omega_\T=C_\T k$ are the corresponding acoustic, \alfven, and tube frequencies, where $k=\pi n/L_0$ is the longitudinal wavenumber, $n$ is the longitudinal mode number ($n=1$ corresponds to the fundamental mode), $L_0$ is the length of the loop, $\gamma=5/3$ is the adiabatic index. 

The boundary value problem (\eqref{eq:cont}-\ref{eq:entropy}) is solved in cylindrical coordinates ($r,\phi,z$) with the Neumann boundary conditions at $r=a$,

\begin{align}
 [P_{\T}]_{r=a}  &=0, \\
 [ \xi_r]_{r=a}  &=0,
\end{align}
and the Dirichlet boundary conditions at $r=0,\infty$
\begin{align}
P_{\T}|_{r=0}  &<\infty, \\
\bvec{\xi}^2|_{r=0}  &<\infty, \\
P_{\T}|_{r\to\infty}  &=0, \\
\bvec{\xi}^2|_{r\to\infty}  &=0,
\end{align}
where $P_{\T}$ and $\xi_r$ are the total pressure and the radial displacement, respectively. 
In the case of the standing kink mode ($m=1$), we Fourier-analyse the perturbed quantities by assuming $P_{\T1}=A\RR(r) \cos (\omega t) \sin (k z)\cos(\phi)$, where $A$ is the amplitude of the perturbed total pressure. The longitudinal profile $\sin (k z)$ ensures that the transverse displacement follows a $\sin k z$-distribution, and therefore has a maximum at the loop apex for the fundamental mode ($n=1$). 

The perturbed total pressure $P_{\T1}$ (and $\RR$) must satisfy 
\begin{equation}
\frac{\d^2P_{\T1}}{\d r^2}+\frac{\d P_{\T1}}{r\d r}-(\kappa_r^2+\frac{1}{r^2})P_{\T1}=0, \label{eq:ptr}
\end{equation} 
where $\kappa_r^2=\frac{(\omega_\s^2-\omega^2)(\omega_\A^2-\omega^2)}{(\omega_\s^2+\omega_\A^2)(\omega_\T^2-\omega^2)}k^2$ is the square of the radial wavenumber and has the dimensionality of wavenumber $k^2$. \eqref{eq:ptr} holds for both internal and external plasma, where all quantities are piecewise functions of $r$, and gives $\RR= J_1(|\kappa_{r\i}| r) $ or $K_1(\kappa_{r\e}r)$ for $r <a$ and $r>a$, respectively, where $J_1$ and $K_1$ are the first order Bessel function of the first kind and the first order modified Bessel function of the second kind, respectively. We re-define $|\kappa_{r\i}|=\sqrt{-\kappa_{r\i}^2}$, so the dispersion relation for the fast body mode is obtained:
\begin{equation}
\frac{\kre}{\rho_\e(\omega_{\A\e}^2-\omega^2)}\frac{K'_1(\kre a)}{K_1(\kre a)}=\frac{\kri}{\rho_\i(\omega_{\A\i}^2-\omega^2)}\frac{J'_1(\kri a)}{J_1(\kri a)}. \label{eq:disp}
\end{equation}

The perturbed Lagrangian quantities used in the Forward Modelling code are:
\begin{align}
 v_r &=\hat{v}_r(r) \sin (\omega t) \sin (k z) \cos\phi, \label{eq:vr} \\
 v_\phi& =\hat{v}_\phi(r) \sin (\omega t) \sin (k z) \sin\phi,  \\
 v_z &=\hat{v}_z(r) \sin (\omega t) \cos (k z) \cos\phi,\\
 \rho_1&=\hat{\rho}_1(r)\cos (\omega t) \sin (k z) \cos\phi, \\
 T_1&= \hat{T}_1(r) \cos (\omega t) \sin (k z) \cos\phi, \label{eq:temp} %(\gamma-1)T_0\frac{\rho_1}{\rho_0}
\end{align}
where 
\begin{align}
 \hat{v}_r&=-\frac{r \d \RR }{\RR\d r}\hat{v}_0(r), \\
 \hat{v}_\phi&=\hat{v}_0(r)= \frac{A\RR\omega}{r\rho_0(\omega^2-\omega_\A^2)}, \\
 \hat{v}_z &=-\frac{C_\T^2 kr}{C_\A^2}\frac{(\omega^2-\omega_\A^2)}{(\omega^2-\omega_\T^2)} \hat{v}_0, \\
 \hat{\rho}_1&=\frac{(\omega^2-\omega_\A^2)}{(\omega^2-\omega_\T^2)} \frac{\rho_0 r \omega \hat{v}_0}{(C_\s^2+C_\A^2)}, \\
 \hat{T}_1&=\frac{(\omega^2-\omega_\A^2)}{(\omega^2-\omega_\T^2)} \frac{(\gamma-1)T_0 r\omega \hat{v}_0}{(C_\s^2+C_\A^2)}.
\end{align}

The horizontally polarised kink mode has
\begin{align}
v_x&=\tilde{v}_x(r,\phi)\sin (\omega t) \sin (k z), \\
v_y&=\tilde{v}_y(r,\phi)\sin (\omega t) \sin (k z),
\end{align}
where
\begin{align}
\tilde{v}_x&=\hat{v}_r\cos^2\phi-\hat{v}_\phi \sin^2\phi \\
      &=\left\{
	  \begin{array}{lr}
	      v_{00}(J_0-J_2\cos2\phi),\;\text{for $r \leq a$} \\
	      -\frac{J'_1(|\kappa_{r\i}|a)}{K'_1(\kappa_{r\e}a)}v_{00}(K_0+K_2\cos 2\phi), \;\text{for $r>a$},
	  \end{array}\right.\label{eq:vx}\\
\tilde{v}_y&=\hat{v}_r\cos\phi\sin\phi +\hat{v}_\phi\cos\phi\sin\phi \\
       &=\left\{
 	  \begin{array}{lr}
 	      -v_{00}J_2\sin2\phi,\;\text{for $r \leq a$} \\
 	      -\frac{J'_1(|\kappa_{r\i}|a)}{K'_1(\kappa_{r\e}a)}v_{00}K_2\sin 2\phi,\;\text{for $r>a$},
 	  \end{array}\right.\label{eq:vy}
\end{align}
and $v_{00}=-\frac{A_\i\omega |\kappa_{r\i}| }{2\rho_\i(\omega^2-\omega^2_{\A\i})}$ is the Lagrangian velocity at the $r=0$.

We could see that the plasma motion is predominantly polarised along the x-direction described by the $J_0$ term  \citep[also see Appendix \ref{sec:appendix} and][]{goossens2014}. The quadrupole terms $J_2\cos(2\phi)$ and $J_2\sin(2\phi)$ may contribute to the fine structuring of coronal loops associated with kink modes. The vertically polarised transverse mode could be easily obtained by replacing $\phi$ with $\phi+\pi/2$, while keeping the coordinate system intact.

We rewrite Equations \ref{eq:vx} and \ref{eq:vy} as
\begin{align}
\begin{bmatrix}
\tilde{v}_x \\
\tilde{v}_y
\end{bmatrix}
&=
\begin{bmatrix}
\tilde{v}_x^{[1]} \\
\tilde{v}_y^{[1]}
\end{bmatrix}
+
\begin{bmatrix}
\tilde{v}_x^{[2]} \\
\tilde{v}_y^{[2]}
\end{bmatrix}, \\
\begin{bmatrix}
\tilde{v}_x^{[1]} \\
\tilde{v}_y^{[1]}
\end{bmatrix}
&=\left\{
\begin{array}{lr}
v_{00}
\begin{bmatrix}
J_0 \\
0
\end{bmatrix} ,\;\text{for $r \leq a$},\\
-\frac{J'_1(|\kappa_{r\i}|a)}{K'_1(\kappa_{r\e}a)}v_{00}
\begin{bmatrix}
K_0 \\
0
\end{bmatrix},\;\text{for $r>a$,}
\end{array}\right. \label{eq:vxy1} \\
\begin{bmatrix}
\tilde{v}_x^{[2]} \\
\tilde{v}_y^{[2]}
\end{bmatrix}
&=\left\{
\begin{array}{lr}
-v_{00}J_2
\begin{bmatrix}
\cos2\phi \\
\sin2\phi
\end{bmatrix} ,\;\text{for $r \leq a$}\\
-\frac{J'_1(|\kappa_{r\i}|a)}{K'_1(\kappa_{r\e}a)}v_{00}K_2
\begin{bmatrix}
\cos2\phi \\
\sin2\phi
\end{bmatrix},\;\text{for $r>a$.}
\end{array}\right. \label{eq:vxy2}
\end{align}

\begin{figure*}[ht]
\centering
\includegraphics[width=0.6\textwidth]{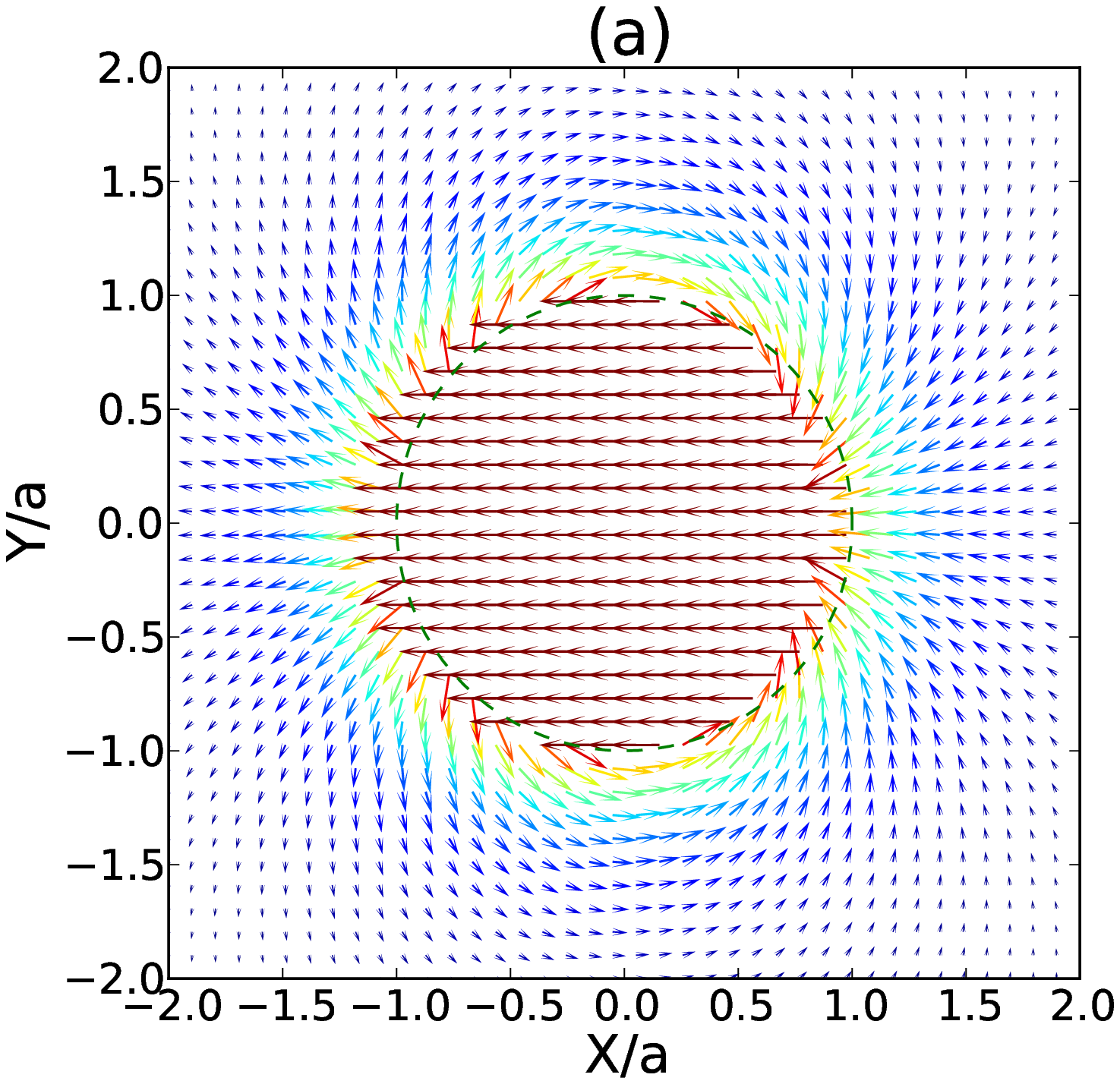}
\includegraphics[width=0.48\textwidth]{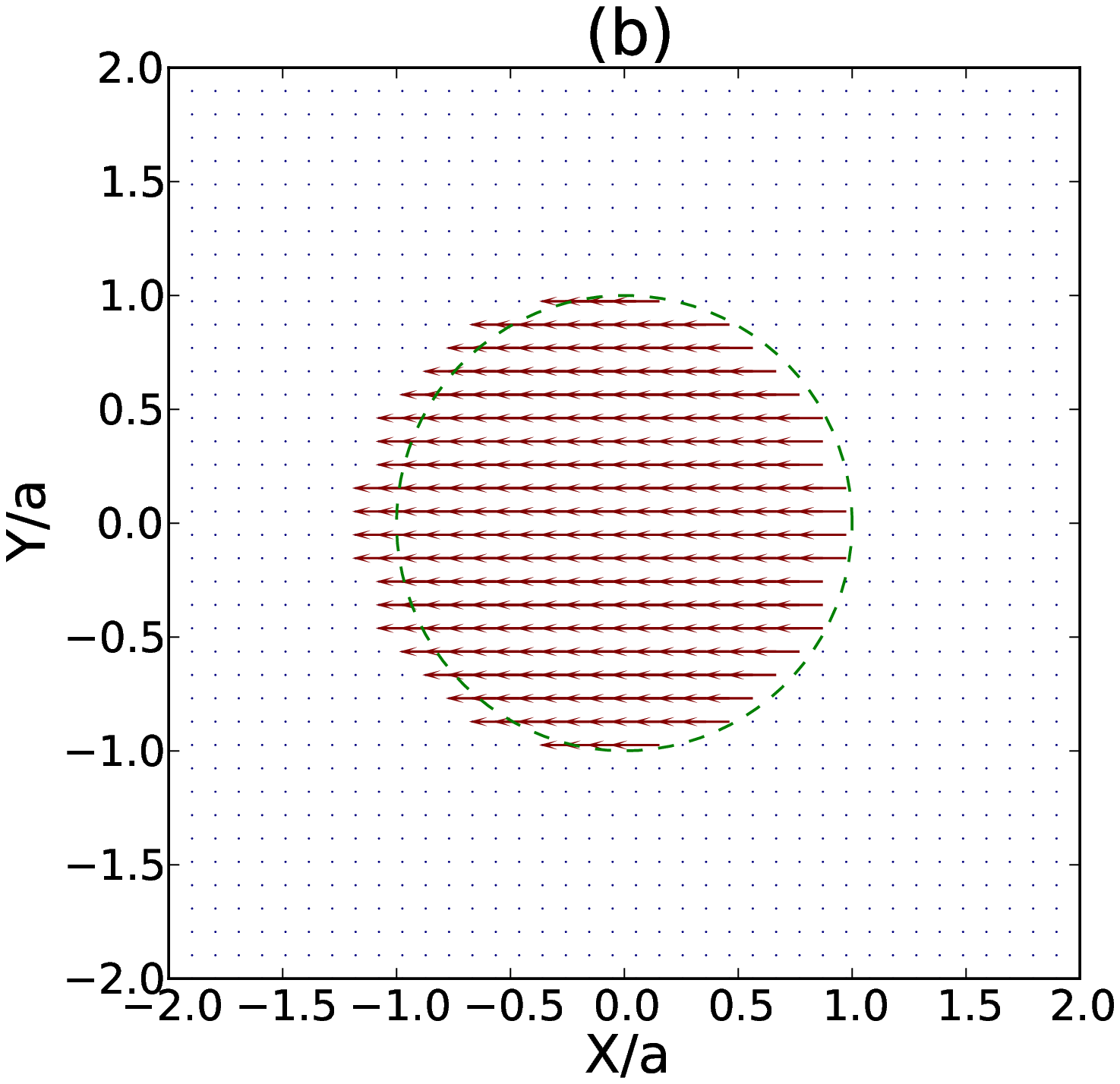}
\includegraphics[width=0.48\textwidth]{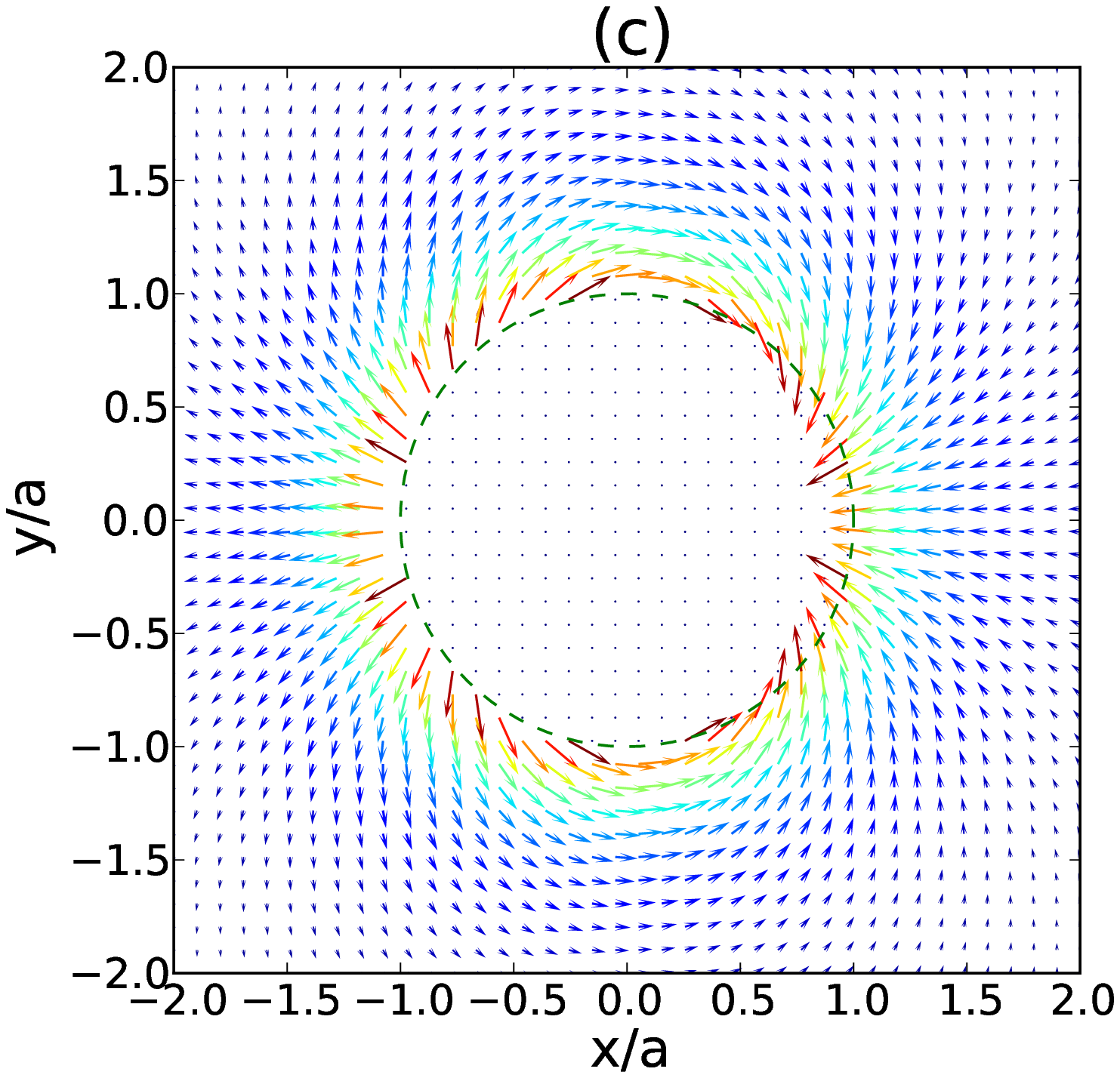}
\caption{Flow fields at cross-section of $z=L_0/2$. (a) illustrates the full solution $[\tilde{v}_x, \tilde{v}_y]^\mathrm{T}$, whereas (b) and (c) shows the polarized quasi-rigid motion $[\tilde{v}_x^{[1]}, \tilde{v}_y^{[1]}]^\mathrm{T}$ and the quadrupole term $[\tilde{v}_x^{[2]}, \tilde{v}_y^{[2]}]^\mathrm{T}$. \label{fig:deform}}
\end{figure*}

In the thin flux tube limit ($ka\ll1$), $J_0=1+\bigO{(ka)^2}$ for $r<a$, and $K_0/K_2\ll1$ for $r>a$, so the polarized quasi-rigid motion is almost confined within the tube $r<a$ (\figref{fig:deform}{b}). The quadrupole term is of secondary effect, as $J_2=\bigO{(ka)^2}$ for $r<a$, whereas at $r>a$, $K_2/K_0\gg1$, so it dominates the surrounding plasma (\figref{fig:deform}{c}). But we shall note that the quadrupole term is only a second order term inside the tube, whereas at the ambient plasma, its magnitude is of the first order.

\subsubsection{Correction for advected plasma motion}
\label{sec:adv} 
Equations~\ref{eq:vr}-\ref{eq:temp} are solutions in Lagrangian coordinates, while we need to synthesize observables at a fixed LOS (Eulerian coordinates), therefore, the Lagrangian variables are remapped into Eulerian coordinates. The transverse displacement of a kink mode is of the order of the loop radius $a$ \citep{aschwanden2002}, and therefore, the advected motion cannot be neglected. The displacement $\bvec{\xi}$ for the plasma fluid at initial position $[r,\phi,z]$ could be obtained by integrating the velocity with respect to time $t$. 
\begin{align}
\xi_r &= -\hat{v}_r/\omega \cos (\omega t) \sin (k z) \cos\phi, \label{eq:xir} \\
\xi_\phi& =-\hat{v}_\phi/\omega \cos (\omega t) \sin (k z) \sin\phi, \label{eq:xiph} \\
\xi_z &= -\hat{v}_z/\omega \cos (\omega t) \cos (k z) \cos\phi.
\end{align}
In Cartesian coordinates, the displacement is given as, 
\begin{align}
\xi_x &= \tilde{\xi}_x\cos (\omega t) \sin (k z),\\
\xi_y& =\tilde{\xi}_y\cos (\omega t) \sin (k z),
\end{align}
where $[\tilde{\xi}_x,\tilde{\xi}_y]^\mathrm{T}=-[\tilde{v}_x,\tilde{v}_y]^\mathrm{T}/\omega$.
Then the new position $[\tilde{x},\tilde{y},\tilde{z}]^\mathrm{T}$ of the plasma fluid originally at $[x,y,z]^\mathrm{T}=[r\cos\phi,r\sin\phi,z]^\mathrm{T}$  is
\begin{align}
\tilde{x}(t)&=x+\xi_x(t), \\
\tilde{y}(t)&=y+\xi_y(t), \\
\tilde{z}(t)&=z+\xi_z(t).
\end{align}
Thus, the plasma properties (e.g. $\rho_0+\rho_1$) at location $[x,y,z]^\mathrm{T}$
will be moved to the position $[\tilde{x},\tilde{y},\tilde{z}]^\mathrm{T}$.

\subsubsection{Mapping into a semi-torus structure} 
\label{sec:torus}

\begin{figure}[ht]
\centering
\includegraphics[width=0.5\textwidth]{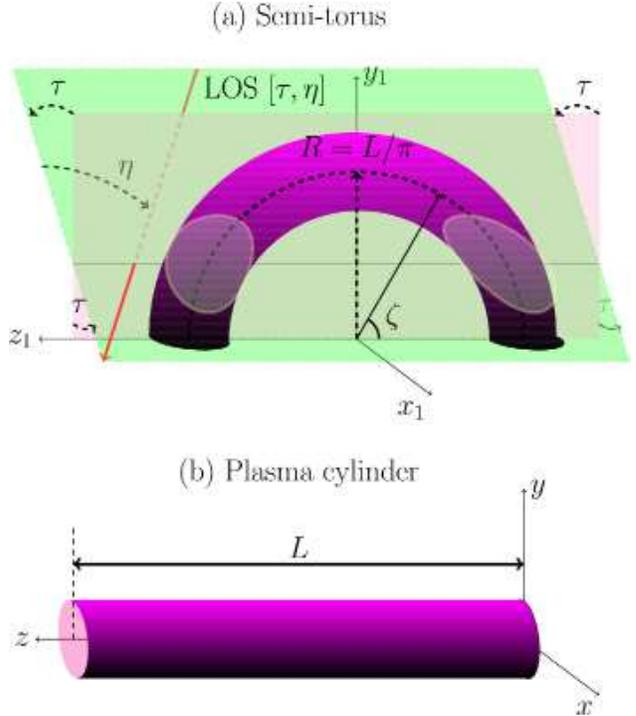} % trim={0 5cm 0 0},clip=true
\caption{Schematic drawing illustrates how the cylinder is mapped into a semi-torus and the LOS angle definition. The pink plane ($y_1z_1$ plane) is defined by the loop spine. The green plane forms an angle of $\tau$ with the pink plane; and their line of intersection is parallel with the $z_1$-axis. The LOS is free to vary within the green plane and quantified by an angle $\eta+\pi/2$ relative to the $z_1$-axis or the line of intersection. A LOS angle is denoted as $[\tau,\eta]$.\label{fig:torus}}
\end{figure}

The loop curvature was found to have a secondary effect on the transverse motion of coronal loops \citep{vandoorsselaere2009}, and therefore we only consider the LOS effect and plasma inhomogeneities by mapping the kink mode solution of a plasma cylinder into a semi-torus structure (see \figref{fig:torus}). The kink mode displaces the axis of the loop, therefore, it has a polarisation relative to the plane defined by the static curved loop axis, i.e., the $y_1z_1$ plane. It is defined as a horizontal kink mode if the loop oscillates out of the $y_1z_1$ plane \citep[e.g.,][]{nakariakov1999,aschwanden1999}. Or otherwise, if the transverse motion of the loop axis is within the $y_1z_1$ plane, it is termed as a vertical kink mode \citep[e.g.,][]{wang2004,verwichte2006}.

After correcting the advected motion, we map the plasma coordinates and the associated plasma parameters into a semi-torus structure (\figref{fig:torus}). The plasma cylinder is bent into a torus within the $y_1z_1$ plane, using the following transform \citep[also see][]{kuznetsov2015}:
\begin{align}
x_1&=x, \\
y_1&=(R+y)\sin\zeta, \\
z_1&=-(R+y)\cos\zeta,
\end{align}
where $\zeta=z/R$, and $R=L/\pi$.

The velocity is transformed by 
\begin{equation}
\begin{bmatrix}
v_{x_1} \\
v_{y_1} \\
v_{z_1}
\end{bmatrix}
=
\begin{bmatrix}
1  & 0          & 0 \\
0  &  \sin\zeta(z) & \cos\zeta(z) \\
0  & -\cos\zeta(z) & \sin\zeta(z)
\end{bmatrix}
\begin{bmatrix}
v_x \\
v_y \\
v_{z}
\end{bmatrix}.
\end{equation}
This is basically a rotation of the velocity vector by an angle of $\zeta-\pi/2$ about the $x$-axis; and $\zeta-\pi/2$ varies within $[-\pi/2,\pi/2]$ for $z\in[0,L]$.

\subsection{Coronal loop model}

Coronal loops are highly complex and dynamic structures, observed within a broad range of plasma conditions, see a review by \citet{reale2014}. The loop width varies from a few hundreds \citep{brooks2013,morton2013} to dozens of thousands kilometres \citep{aschwanden2002,aschwanden2005,aschwanden2011}. A coronal loop may have multi-thermal \citep{nistico2014b}, multi-stranded structures \citep{scullion2014,peter2013}, and may be associated with heating and flows \citep{klimchuk2006,hood2009,winebarger2002}. In our study, these fine structures are not considered; and the gravitational stratification is also neglected. Numerical simulations are required to  model these features of coronal loop oscillations.

A coronal loop is set up in an equilibrium state. The loop measures $L_0=100\unit{Mm}$ in length and $a=4\unit{Mm}$ in radius. The loop density and temperature are $\rho_\i=2.5\cdot10^{-12}\unit{kg\cdot m^{-3}}$ ($n_{e\i}=1.5\cdot 10^9\unit{cm^{-3}}$) and $T_i=0.8\unit{MK}$, respectively. The internal plasma is permeated by a uniform magnetic field $B_\i=15\unit{G}$. We choose a density and temperature ratio of $\rho_\i/\rho_\e=5$ and $T_\i/T_\e=1.5$, respectively, then the magnetic field strength ratio is obtained by balancing the total pressure at the loop boundary. The plasma beta gives $\beta_\i=0.037$ and $\beta_\e=0.0048$ for the internal and external plasma, respectively. The corresponding acoustic speeds are $C_{\s\i}=150\unit{km\,s^{-1}}$ and  $C_{\s\e}=120\unit{km\,s^{-1}}$, while the \alfven speeds are $C_{\A\i}=840\unit{km\,s^{-1}}$ and  $C_{\A\e}=1900\unit{km\,s^{-1}}$. These parameters are commonly observed in coronal loops \citep[e.g.,][]{reale2014,aschwanden2011b}.

For the fundamental mode ($n=1$), the wavelength is much longer than the loop radius ($ka=0.13$). The dispersion relationship (\eqref{eq:disp}) finds a kink mode solution with a period at $P_0=3.0\unit{min}$ ($\omega=0.034$). We choose $A_i=0.15\unit{Pa}$, so that the velocity perturbation amplitude is about $55\unit{km\,\s^{-1}}$, and the amplitude of displacement about $1.6\unit{Mm}$ ($0.4a$). The kink mode could be considered as highly incompressible \citep{vandoorsselaere2008,goossens2012}, the density (temperature) perturbation is about 0.4\% (0.3\%) of the equilibrium value. These parameters are commonly observed by the TRACE and SDO/AIA instruments \citep{aschwanden2002,aschwanden2011}.

\subsection{Forward model}
The loop system was discretised as given by Equations \ref{eq:vr}-\ref{eq:temp} in Cartesian coordinates. We calculate the plasma properties in a domain of $x(y)\in[-2a,2a]$ and $z\in[0,L_0]$ with $160\times160\times400$ grid cells. Forward modelling was performed with a fixed output mesh grid $N_{x_2} \times N_{y_2}=170\times340$\footnote{We refer to the projected output plane as $x_2y_2$ plane.} \citep[see details in][]{yuan2015fm}.  In contrast to compressive MHD modes \citep{antolin2013,reznikova2014,reznikova2015,kuznetsov2015,yuan2015fm}, the kink mode only perturbs the density and temperature to the order of $10^{-3}$-$10^{-4}$ of the equilibrium values, therefore the effect of the contribution function is of secondary order. The spatial distributions of the plasma properties play a key role in determining the observational features. So we only present the synthetic emission of the \feix line, however, the results should be applicable to other optically thin lines. The \ion{Fe}{12} $\lambda193.509$ \AA{} line and the AIA 171 and 193 \AA{} channel were also synthesised, but they only produce redundant results.

\begin{figure*}[ht]
\centering
\includegraphics[width=0.8\textwidth]{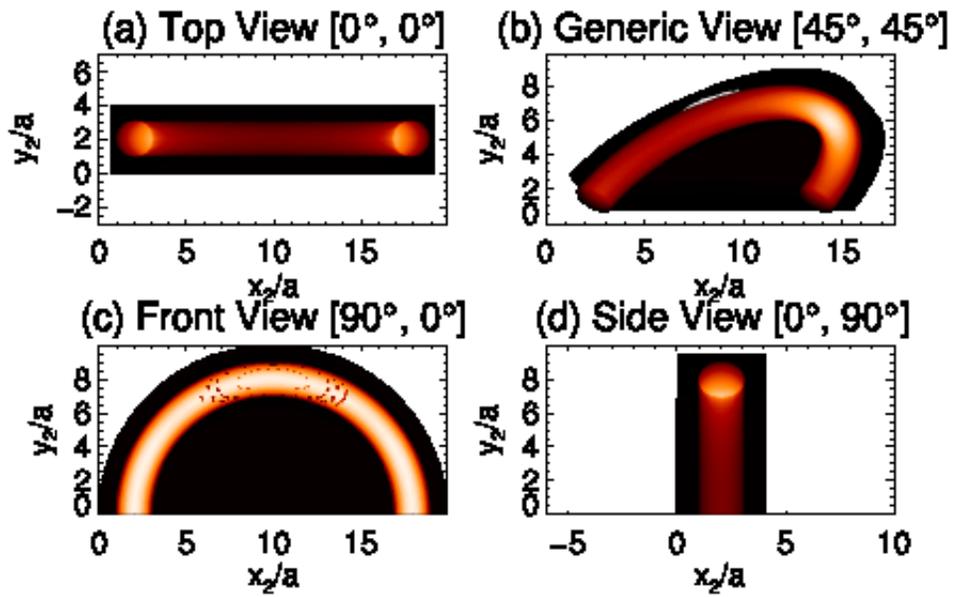} 
\caption{\feix synthetic emission at (a) top view, (b) generic view, (c) front view and (d) side view. The origins and ranges of the plots are chosen to best match the relative geometries at various views, and will not affect the results at all. (This figure is also available as an mpeg animation in the electronic edition of the Astrophysical Journal) \label{fig:synthview}}
\end{figure*}

\begin{figure*}[ht]
\centering
\includegraphics[width=\textwidth]{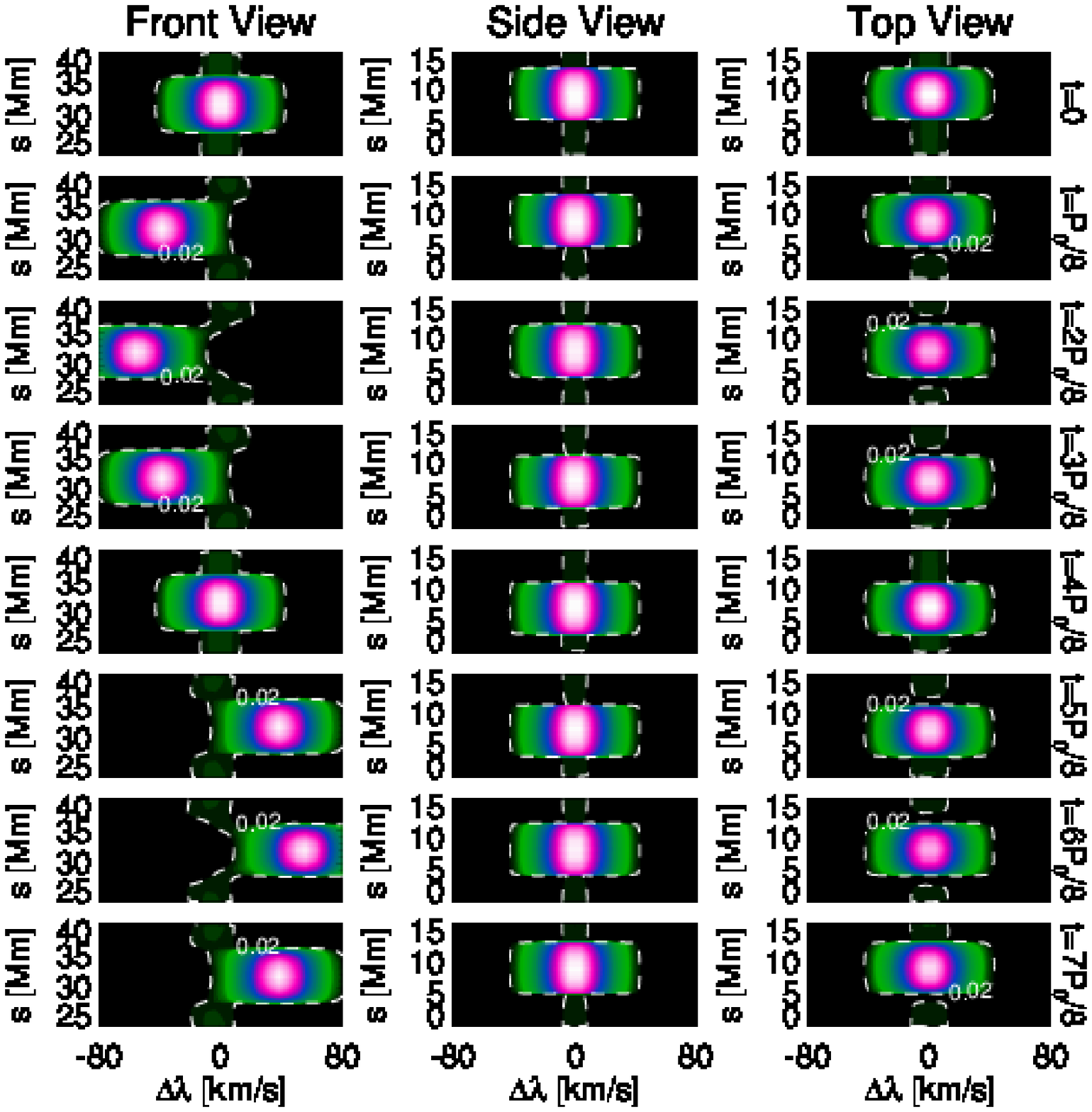}
\caption{\feix spectra of sit-and-stare modes along slices $s$ across the loop apex at front view (left), side view (middle) and top view (right). The dashed lines contour the 0.02-level of the peak emission, and highlight the pendular motion at front view and rotational motion at top view.
(This figure is also available as an mpeg animation in the electronic edition of the Astrophysical Journal)
\label{fig:lambda_yt}}
\end{figure*}

The LOS is defined with two independent angles $[\tau,\eta]$ (see illustration in \figref{fig:torus}), where $\tau$ is the angle between the loop axis plane (pink plane or $y_1z_1$-plane) and another plane (green plane), which share a line of intersection parallel to the $z_1$-axis. The LOS forms an angle of $\eta+\pi/2$ relative to the $z_1$-axis (or the line of intersection). Hereafter, we name $[0\deg,0\deg]$ as top view, $[0\deg,90\deg]$ as side view, $[90\deg,0\deg]$ as front view, and $[45\deg,0\deg]$ as oblique view for reference. \figref{fig:synthview} illustrates the synthetic views in the \feix line at selected viewing angles; while \figref{fig:lambda_yt} presents the spectra of the \feix line along a slice $s$ perpendicular to the axis of the loop apex at each viewing angle, which is comparable with Figure 9 in \citet{goossens2014}. We note that the loop cross-sectional profile could by approximated by integrating the emissivity along a uniform media $2\sqrt{a^2-r^2}$, thus gives a non-Gaussian profile. However, a Gaussian profile is normally assumed and practically observed, e.g., \citet{verwichte2005,aschwanden2011b}. It implies that coronal loops could be multi-thermal \citep[e.g.,][]{nistico2014b}, multi-stranded \citep{peter2013} or inhomogeneous \citep{vandoorsselaere2004}. However, the point spread function may also play a role, especially in low resolution instruments. Inhomogeneity in a coronal loop is favoured by the resonant absorption theory \citep{ruderman2002,vandoorsselaere2004,okamoto2015,antolin2015}, which is developed to explain the strong damping of kink waves \citep{nakariakov1999}. In this study, we do not consider the resonant absorption layer.

\section{Results}
\label{sec:result}

\subsection{Top view}
\label{sec:top}
\begin{figure*}[ht]
\centering
\includegraphics[width=0.8\textwidth]{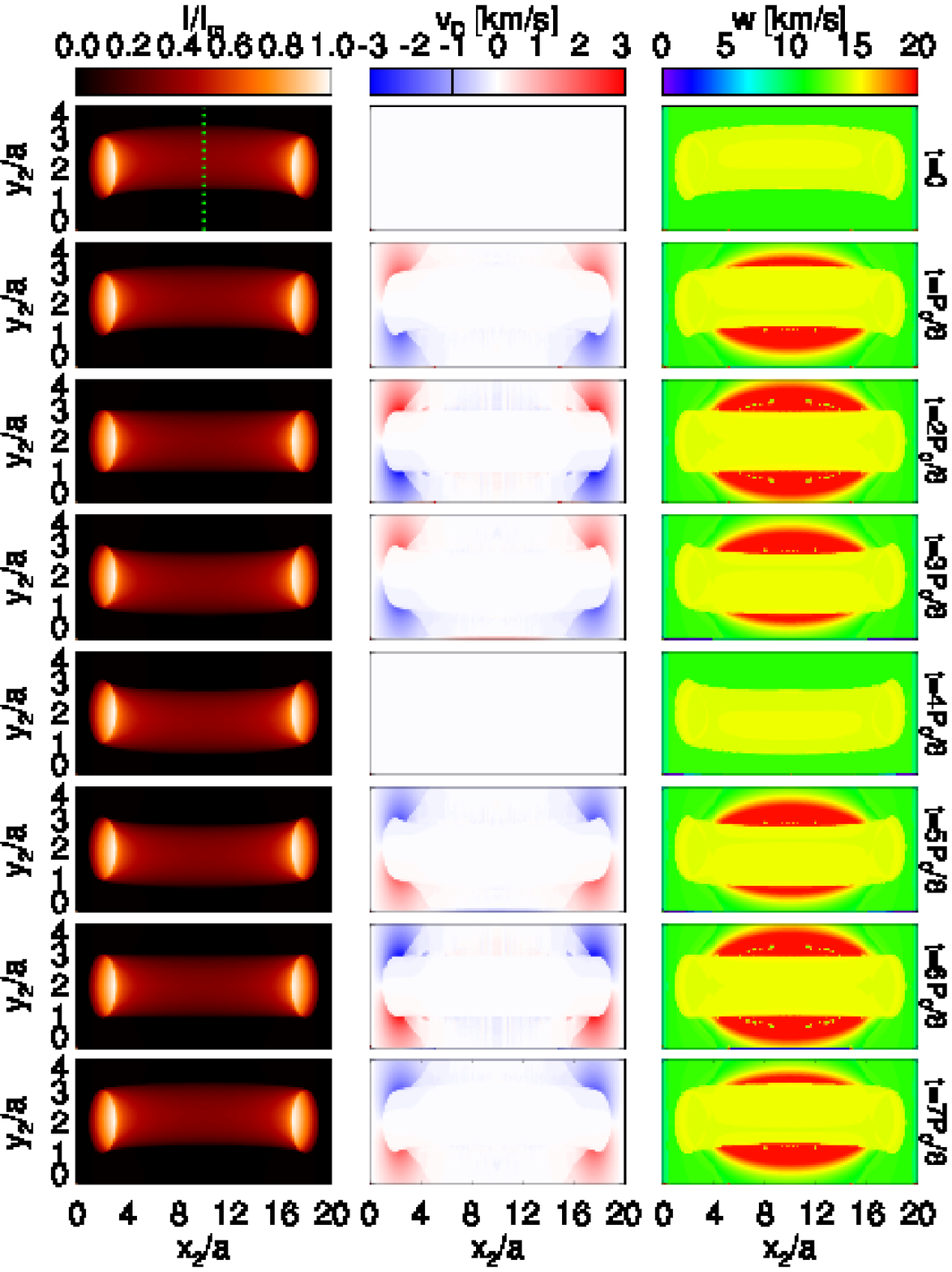}
\caption{Snapshots of the relative emission $I/I_m$ (left), Doppler shift velocity $v_D$ (middle) and line width $w$ (right). (This figure is also available as an mpeg animation in the electronic edition of the Astrophysical Journal)
 \label{fig:topview_fe171}}
\end{figure*}

\figref{fig:topview_fe171} presents snapshots of the relative emission intensity $I/I_m$, Doppler shift velocity $v_D$ and line width $w$ at top view, where $I_m$ is the maximum intensity of the synthetic image series in each viewing angle. At top view, the loop oscillates within the plane-of-sky, it is clearly seen in the relative intensity, Doppler shift velocity, and line width snapshots. The loop motion is not effectively observed in the Doppler shift, as the plasma motion inside and outside the loop is perpendicular to the LOS. 

\begin{figure*}[ht]
\centering
\includegraphics[width=0.8\textwidth]{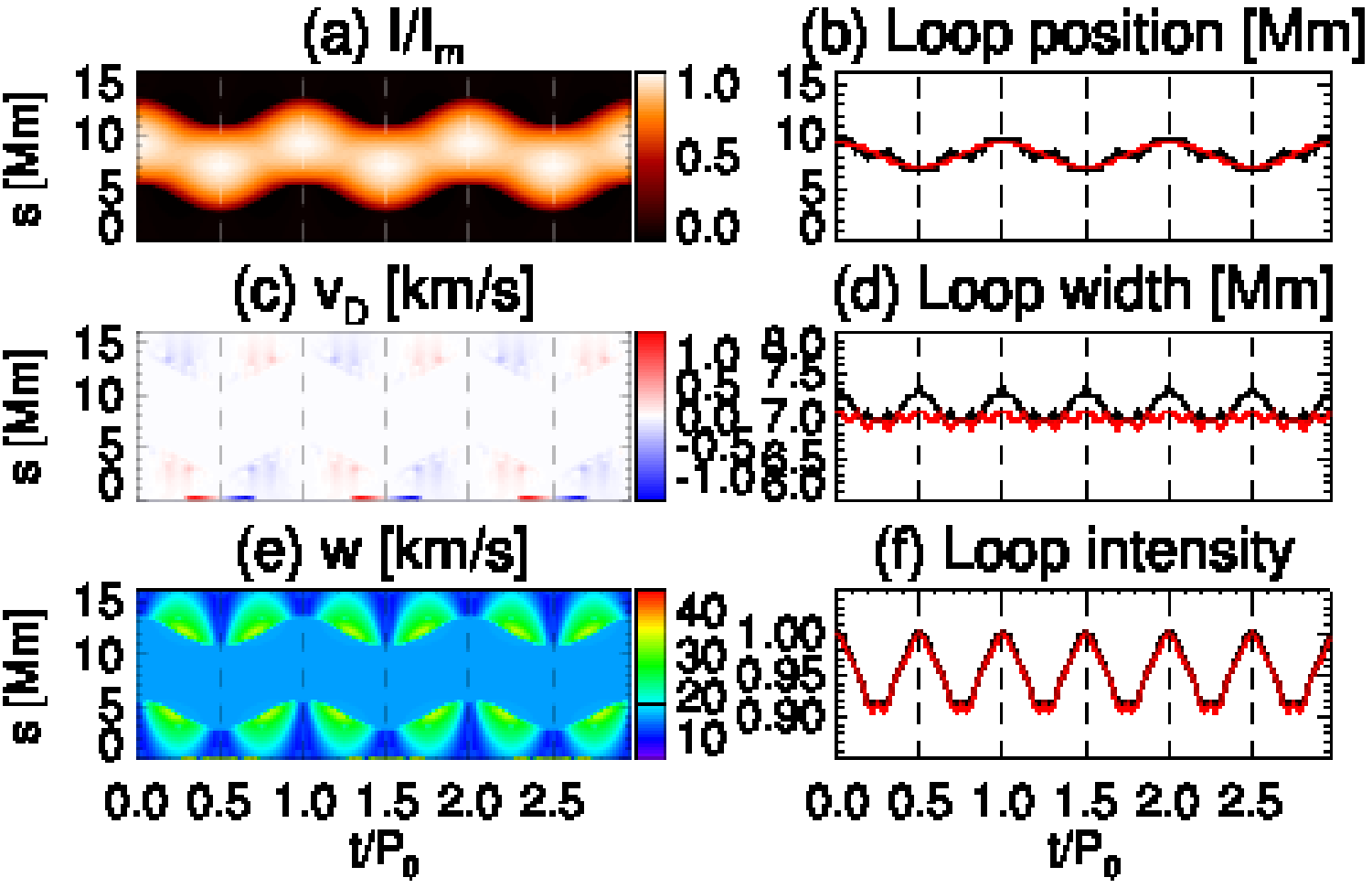}
\caption{Sit-and-stare mode across the loop apex at top view in the \feix line and the measurements of loop position, width and intensity. The red continuous lines plot the corresponding case in the EIS resolution.\label{fig:top_fe171}}
\end{figure*}

\figref{fig:top_fe171} shows the sit-and-stare mode of a spectrograph, e.g., Hinode/EIS, in the \feix line. The time distance plot \citep[see, e.g.,][]{yuan2012sm} is taken at a cut perpendicular to the loop axis, see \figref{fig:topview_fe171}. It clearly shows the transverse loop motion with an amplitude of about $1.5\unit{Mm}$ ($0.4a$); the associated intensity modulation is about 0.04. Intriguingly, we also detect loop width (Full-Width at Half-Maximum) variation between $7.0\unit{Mm}$ ($1.7a$) to $7.4\unit{Mm}$ ($1.8a$); the amplitude is about $0.2\unit{Mm}$, one order of magnitude smaller than the loop displacement. The periodicity of the loop width and intensity variation is $P_0/2$. The Doppler shift velocity $v_D$ is close to zero both inside and outside the loop. Moreover, one could also observe line broadening on the loop periphery. The Doppler shift has the same periodicity $P_0$, but $\pi/4$ out of phase with the transverse motion. The line width variation has a period of $P_0/2$, and oscillates with a phase of $\pi/2$ lagging behind the loop intensity variation. \figref{fig:lambda_yt} (right column) illustrates this effect: the spectrum moves as a whole in space due to the transverse motion, however, the centroid of the spectrum remains unchanged, i.e., $v_D\simeq0\unit{km\,s}$. At the loop periphery, one could observe significant periodic broadening. The background emission, about 2\% of the loop emission, is associated with the quadrupole terms in Equations \ref{eq:vx} and \ref{eq:vy} (see Appendix \ref{sec:appendix} for derivations).

\subsection{Front view}
\begin{figure*}[ht]
\centering
\includegraphics[width=0.8\textwidth]{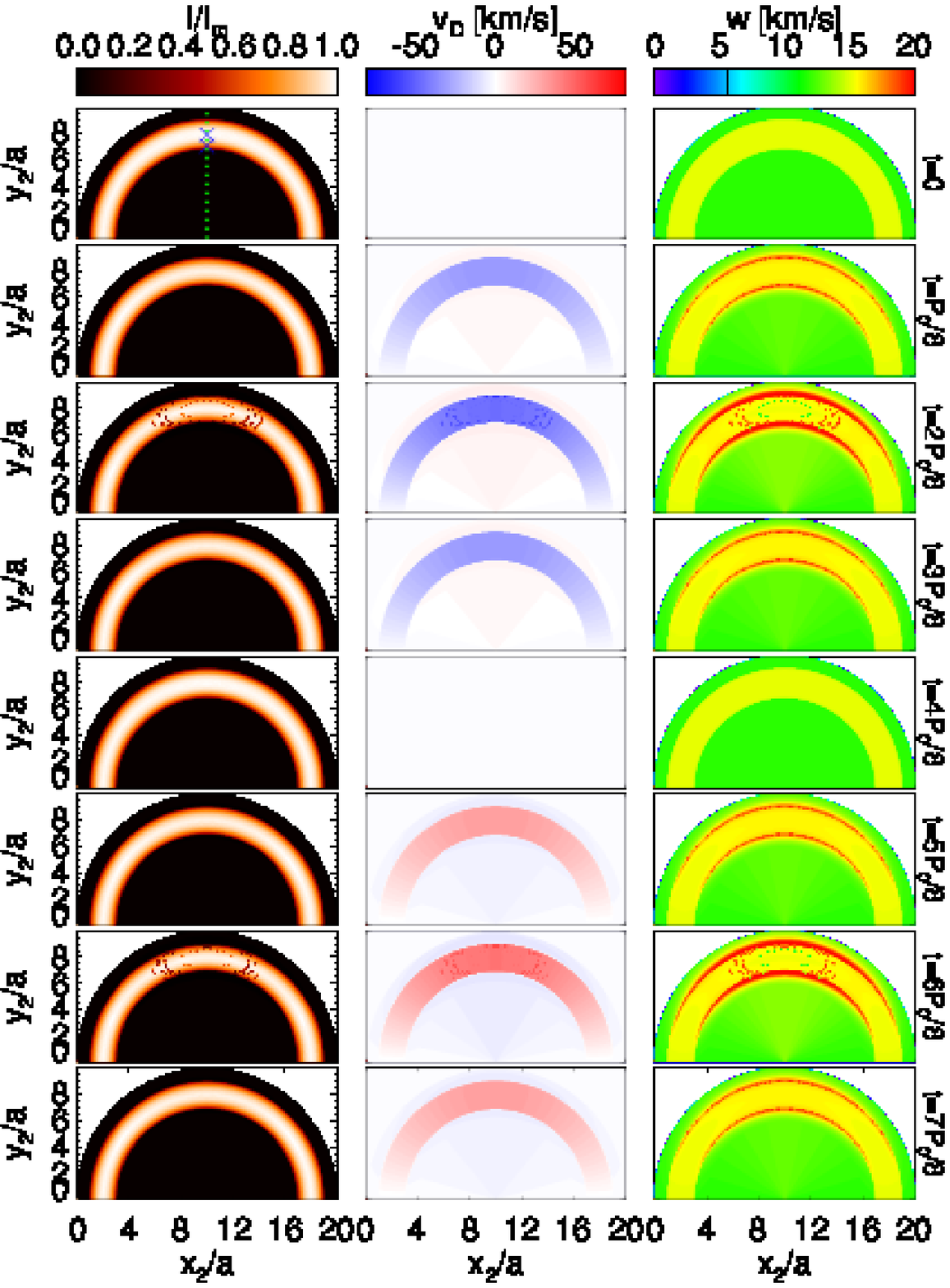}
\caption{The same as \figref{fig:topview_fe171} but at front view. Two crosses mark the positions at $r=0$ and $r=0.8a$ and the associated dynamic spectra are illustrated in \figref{fig:spec}. (This figure is also available as an mpeg animation in the electronic edition of the Astrophysical Journal)\label{fig:frontview_fe171}}
\end{figure*}

\begin{figure*}[ht]
\centering
\includegraphics[width=0.8\textwidth]{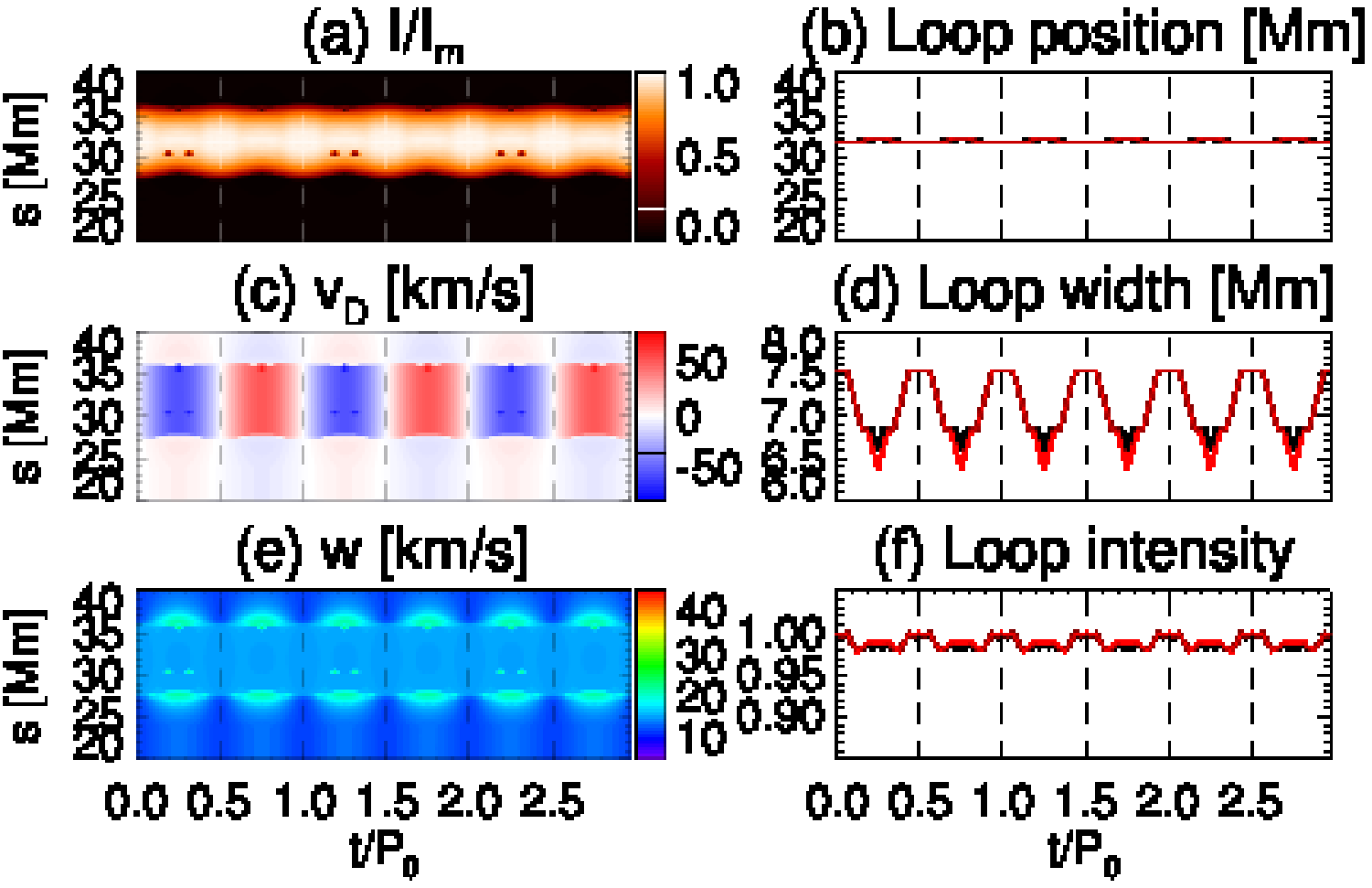}
\caption{The same as \figref{fig:top_fe171} but at front view.\label{fig:front_fe171}}
\end{figure*}

At front view, the transverse motion is along the LOS, so the measured loop displacement is almost zero (\figref{fig:frontview_fe171}). The Doppler shift velocity and line width broadening at the loop edges are detectable. The time-distance plot (\figref{fig:front_fe171}) shows that the loop width oscillates with an amplitude of $0.1a$ and a period of $P_0/2$. The amplitude (about $0.1a$) observed at front view is about twice that (about $0.05a$) measured at top view. Again, we detect line width broadening at the periphery of the loop. This effect may contribute to the non-thermal broadening that has been observed at the edge of active region loops \citep{doschek2007}. The associated Doppler shift (about $5\unit{km\,s^{-1}}$) of the ambient plasma still exists (\figref{fig:front_fe171}), however, in contrast to the apparent rotational motion at top view, the oscillation resembles a pendular motion relative to the loop oscillation. \figref{fig:lambda_yt} (left column) illustrates this effect: the Doppler shift on the periphery of the loop oscillates in anti-phase with the kink motion inside the loop, but with an amplitude of about 10\% of the loop oscillation. This is consistent with Figure~9 in \citet{goossens2014}.

\subsection{Oblique view}
\begin{figure*}[ht]
\centering
\includegraphics[width=0.8\textwidth]{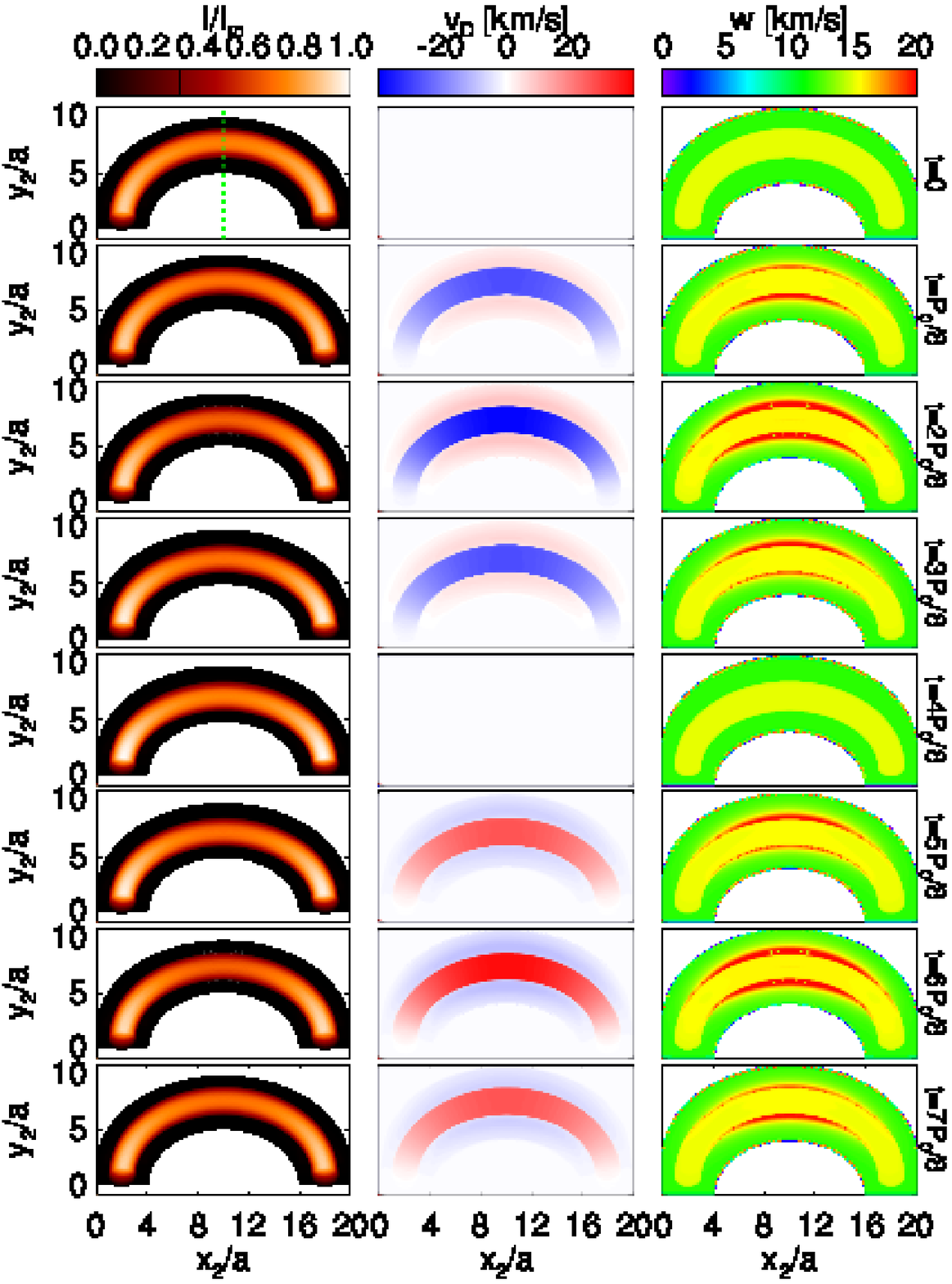}
\caption{The same as \figref{fig:topview_fe171} but at oblique view. (This figure is also available as an mpeg animation in the electronic edition of the Astrophysical Journal)\label{fig:obliqueview_fe171}}
\end{figure*}

\begin{figure*}[ht]
\centering
\includegraphics[width=0.8\textwidth]{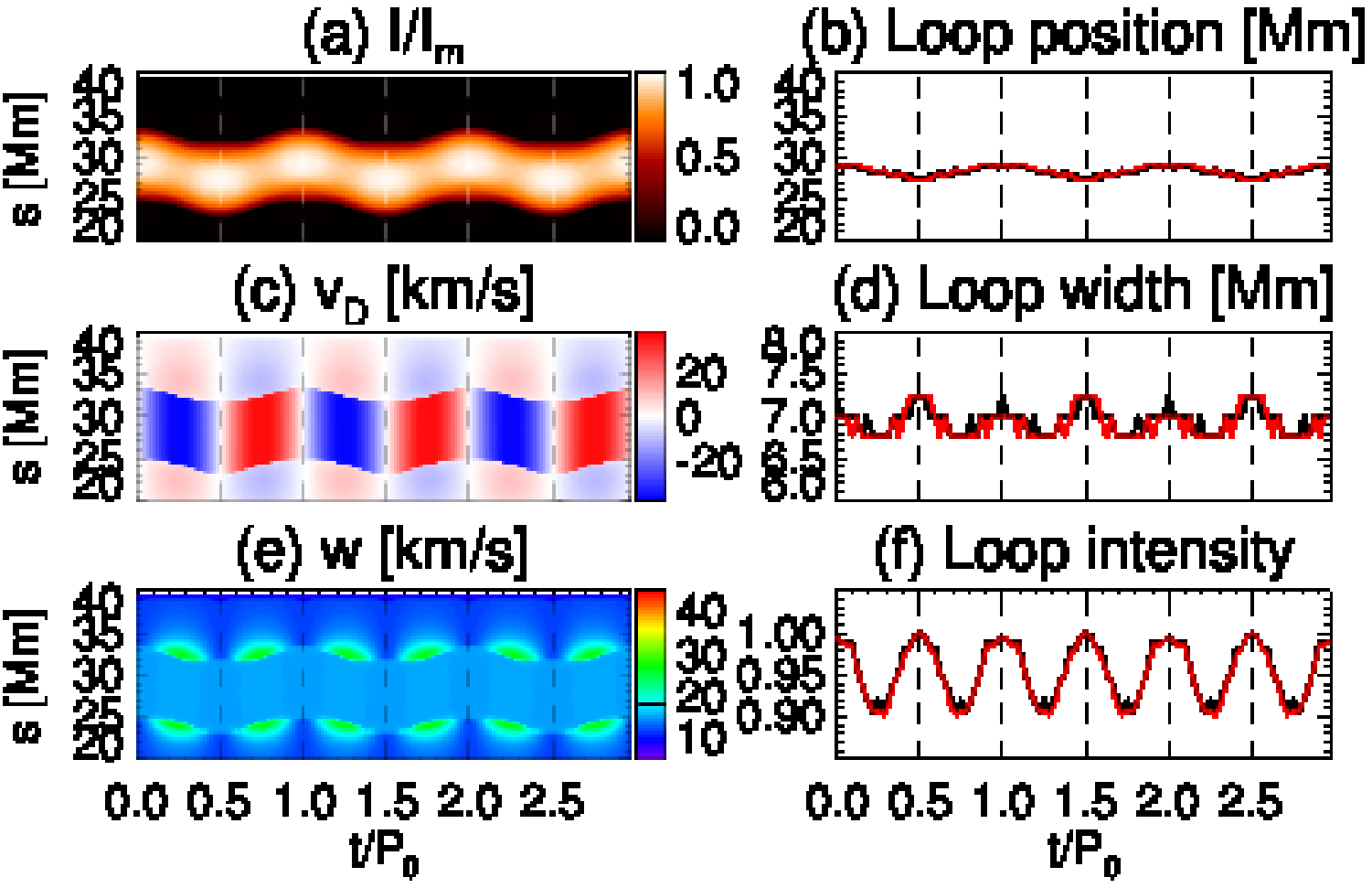}
\caption{The same as \figref{fig:top_fe171} but at oblique view.\label{fig:oblique_fe171}}
\end{figure*}

Oblique view (\figref{fig:obliqueview_fe171}) is the most frequently encountered observation on the solar disk. \figref{fig:oblique_fe171} shows the time-distance plot observed in the \feix line. Loop oscillation features at oblique view contain a mixture of the properties observed at top and front views: loop displacement, intensity modulation and loop width vary at moderate levels.
 
\subsection{Side view}

\begin{figure*}[ht]
\centering
\includegraphics[width=0.8\textwidth]{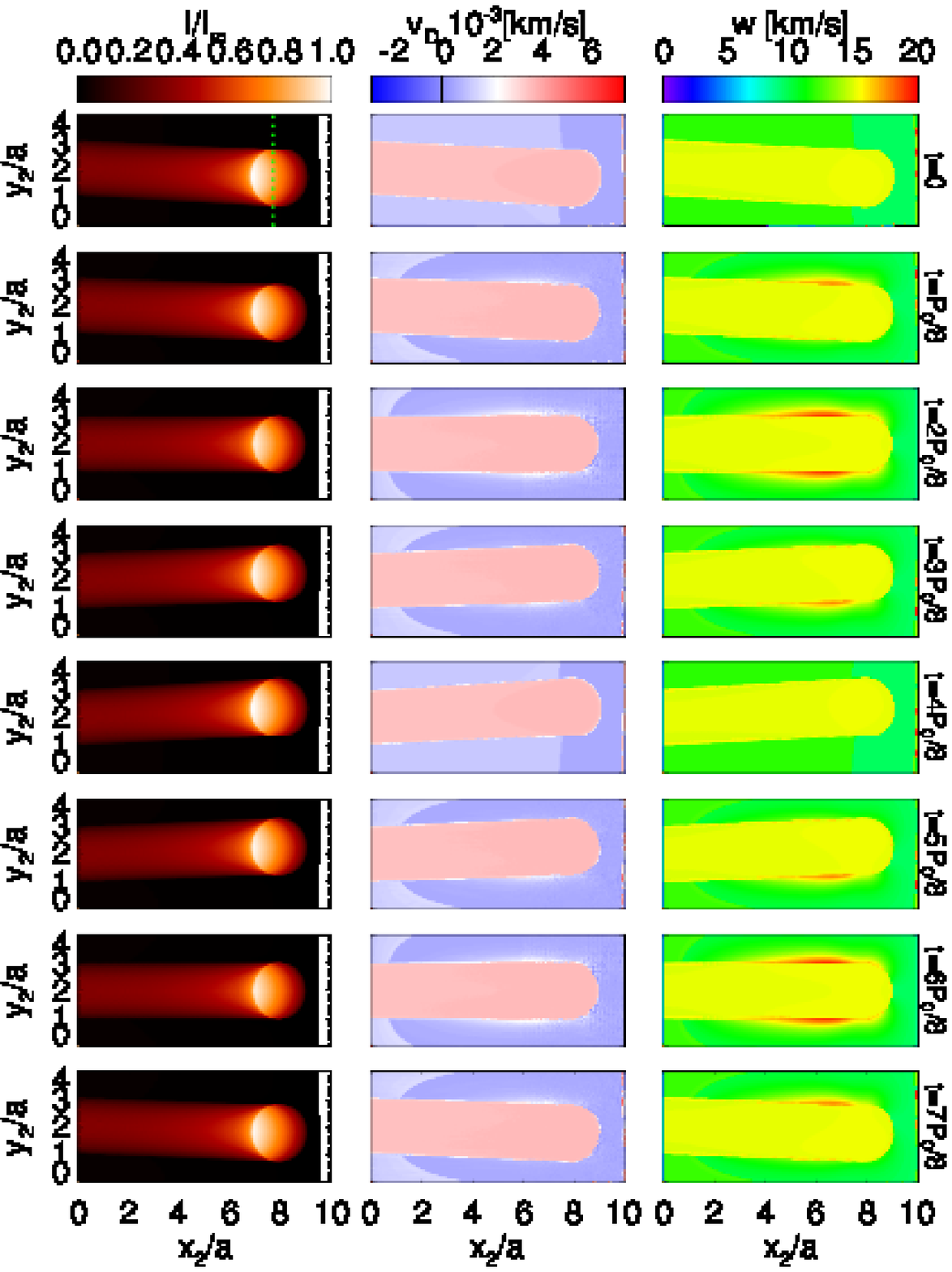}
\caption{The same as \figref{fig:topview_fe171} but at side view. (This figure is also available as an mpeg animation in the electronic edition of the Astrophysical Journal) \label{fig:sideview_fe171}}
\end{figure*}

Side view and its varieties are the most probable viewing angles for off-limb coronal loops, see,  e.g., \citet{verwichte2004}. \figref{fig:sideview_fe171} displays a complete cycle of the standing kink wave at side view. Loop displacement is optimal for observation in the intensity; while the Doppler shift is very small. The line width does not exhibit significant spatial variation over the projected loop. However, line broadening is significantly measurable. The maximum line width broadening is not located at the loop apex, this is because at the apex the plasma motion is almost perpendicular, rather than along the LOS, and the projected fluid motion is only significant at some distance away from the apex.

\begin{figure*}[ht]
\centering 
\includegraphics[width=0.8\textwidth]{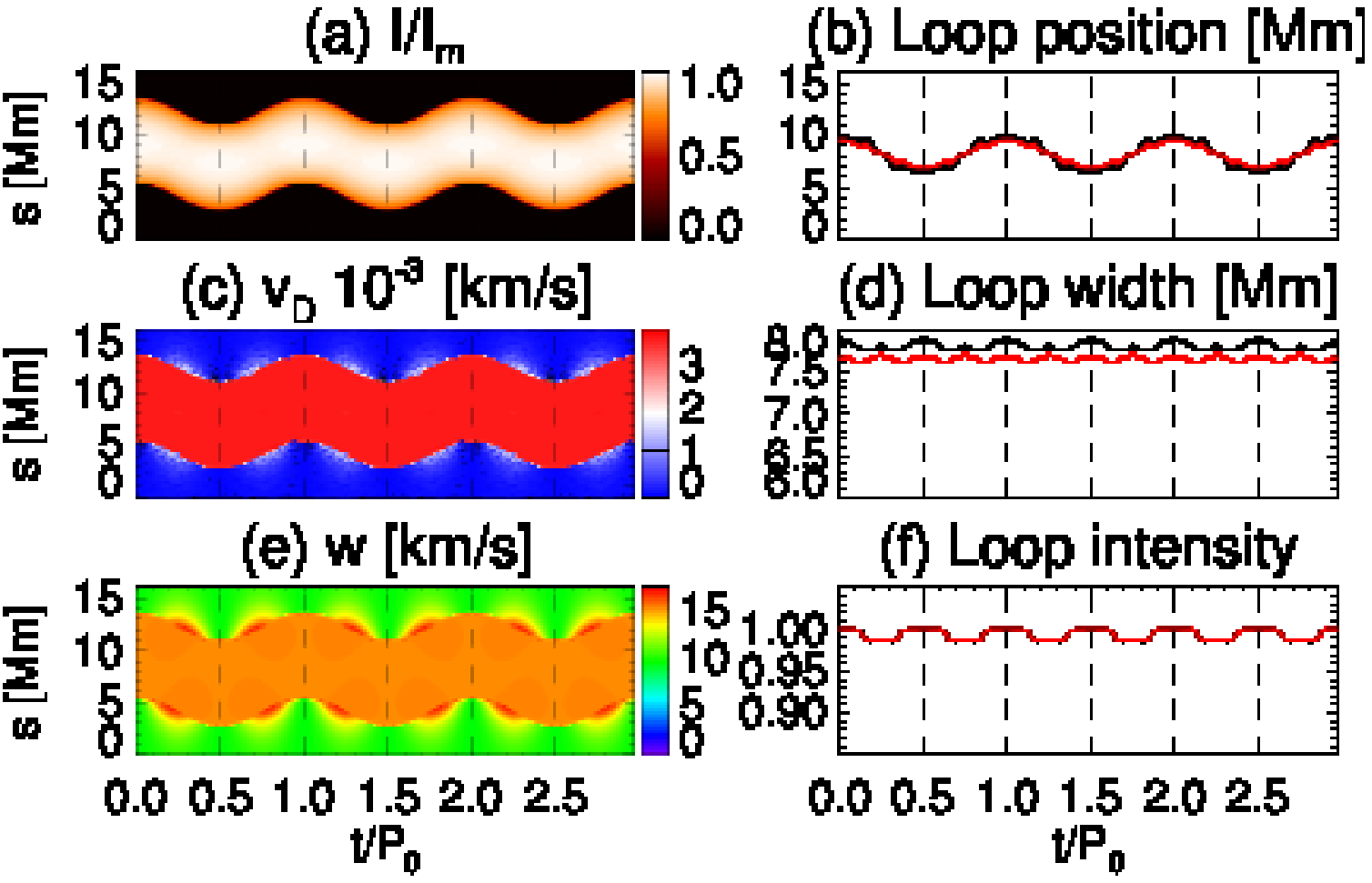}
\caption{The same as \figref{fig:top_fe171} but at side view.\label{fig:side_fe171}}
\end{figure*}

\figref{fig:side_fe171} presents the time-distance plot at the loop apex and the time series of the loop position, width and intensity variations. The times series of the transverse motion is close to a sinusoidal profile, while in other viewing angles the loop displacement deviates significantly from a harmonic function. We note that at side view, the loop width measures at $\simeq8\unit{Mm}$ (about $2a$), whereas other viewing angles normally do not reveal the full width of the loop. The associated loop width and intensity variations are very small.

\begin{figure}[ht]
\centering 
\includegraphics[width=0.5\textwidth]{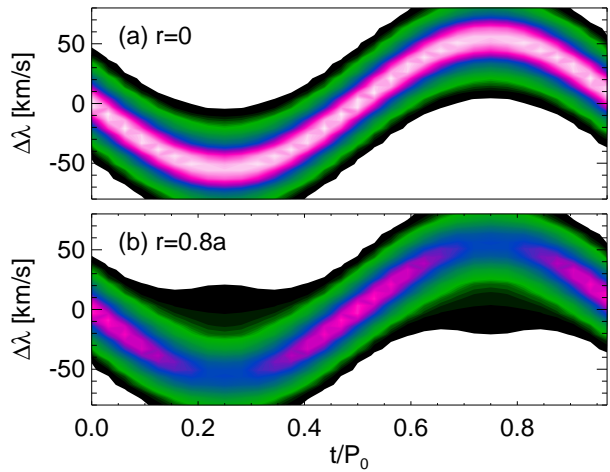}
\caption{Dynamic spectra extracted at positions $r=0$ (a) and $r=0.8a$ (b) at front view, as labeled in \figref{fig:frontview_fe171}.\label{fig:spec}}
\end{figure}

\section{Discussion and Conclusion}
\label{sec:conclusion}
In this study, we discretised the fundamental standing kink wave solution of a plasma cylinder, corrected for the fluid advection, and mapped the solution into a semi-torus structure to simulate the kink MHD mode of a curved coronal loop. Then, we synthesised the EUV emission in the \feix line and performed Gaussian fits to the spectra to obtain the observables, i.e., the emission intensity, Doppler shift velocity and line width.

We find that the cross-sectional intensity distribution of a coronal loop filled with uniform plasma does not follow a Gaussian profile. It means complex coronal loop structure has to be considered to fully synthesize loop oscillations. More physics is associated with loop inhomogeneities, i.e., resonant absorption \citep{ruderman2002,vandoorsselaere2004}, phase mixing \citep{heyvaerts1983}, mode conversion \citep{pascoe2010,pascoe2011,pascoe2012}. 

Loop displacement could be observed in any viewing angles, as long as the polarised motion is not along the LOS. This is the intrinsic feature of a kink MHD wave.

\begin{figure*}[ht]
\centering
\includegraphics[width=\textwidth]{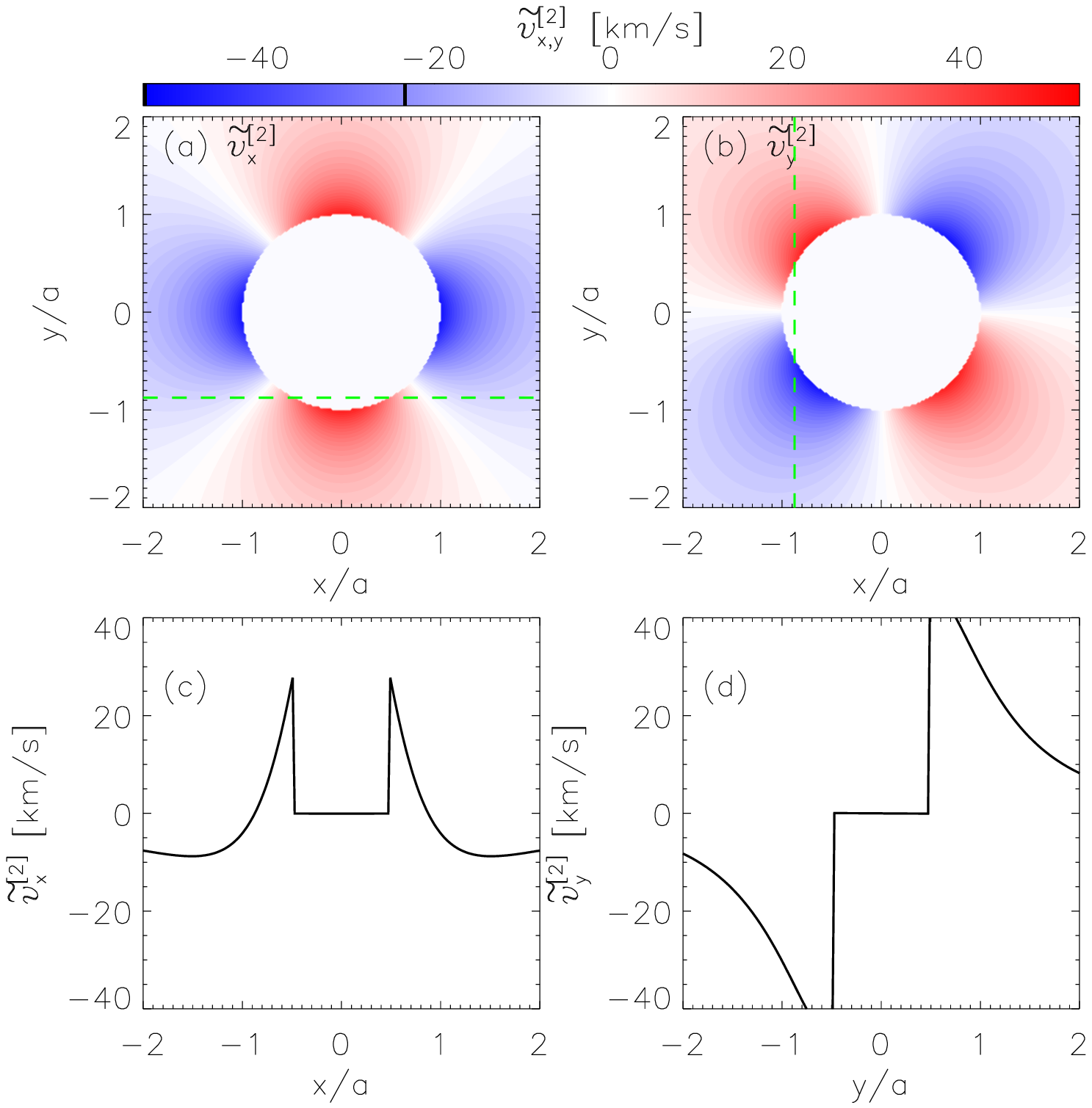}
\caption{(a) and (b) Cross-sectional distribution of the quadrupole terms $\tilde{v}_x^{[2]}$ and $\tilde{v}_y^{[2]}$. (c) Profiles of $\tilde{v}_x^{[2]}$ along the horizontal dashed line marked in (a). (d) Profiles of $\tilde{v}_y^{[2]}$ along the vertical dashed line labeled in (b).\label{fig:cross_xy}}   
\end{figure*}

Since the density and temperature perturbations are of the order of $10^{-3}$-$10^{-4}$ of the equilibrium values, the contribution function has a negligible effect on the loop intensity modulation. The kink mode solution could be decomposed into a quasi-rigid transverse motion and a quadrupole term. The quadrupole term appears in both the $v_x$ and $v_y$ components of the transverse velocity (Equations \ref{eq:vx} and \ref{eq:vy}). \textbf{The fluid elements at $[r,\phi]$ and $[r,-\phi]$ (or equally $[r,\phi]$ and $[r,\pi-\phi]$) outside the tube would periodically deform the $r$-shell in the Lagrangian coordinate at the order of $2\frac{J'_1(|\kappa_{r\i}|a)}{K'_1(\kappa_{r\e}a)}v_{00}K_2/\omega$ (see Equations \ref{eq:vxy1} and \ref{eq:vxy2})}, which is a few percent of the loop radius $a$, if the amplitude of the displacement is close to $a$. Moreover, the fluid elements at $\pm\phi$ of the $r$-shell move to the opposite directions (\figref{fig:cross_xy}), and thus cause spectral line broadening.  The broadening is also accompanied by intensity suppression as illustrated in \figref{fig:spec}. At front view, the emission suppression at $r=0.8a$ is stronger than that at the loop axis. The quadrupole term effect only becomes significant at loop edges, where the LOS integrates through more ambient plasma, and has smaller impact on the spectrum at the loop axis as the major contribution are from the plasma inside the tube.

Line width broadening is usually measured in the periphery of the loop, where ambient plasma emission is significant. It is associated with the $\cos(2\theta)$ and $\sin(2\theta)$ terms in \eqref{eq:vx} and \ref{eq:vy}. The line broadening is observed at all views. \figref{fig:cross_xy} illustrates the reason: at front view $v_x$ could vary from positive to negative along a LOS at loop edge; whereas at top view, $v_y$ is anti-symmetric about the $xz$-plane. This is consistent with the case of a vertical transverse wave \citep{vandoorsselaere2008b}. \citet{doschek2007} reported non-thermal broadening at the edge of active region loops, and it may be connected with kink mode perturbations in the loops. However, since there is no report of the associated transverse motion, it may imply that coronal loops have unresolved low-amplitude motion similar to \citet{nistico2013} and \citet{anfinogentov2013}.

The intensity modulation at the loop axis is usually detectable. This is different with \citet{cooper2003a,cooper2003b}, which only consider static plasma emission. In our study, both the spectrum modification by the MHD wave motion are considered and measured as what would occur in realistic observations. This factor could contribute to the integrated LOS intensity variation at the loop axis.

At front view and its varieties, a pendular motion is observed. At front view, the transverse motion of the loop could be fully observed along the LOS, while the background emission oscillates at an amplitude of a few percent of the loop oscillation amplitude.

It is intriguing that \textbf{coronal loop is observed to have an apparent periodic expansion and contraction} when undergoing a kink MHD mode wave. The optimal viewing angle to observe this effect is the front view. The amplitude of the loop width variation is about 20-30\% of the transverse loop motion. At top view, the loop width variation is about half of the amplitude measured at front view. At side view, this effect could not be observed. The loop deformation introduced by the quadrupole terms alone is not fully responsible for the loop width variation at such amplitude. The line width broadening would result in emission intensity suppression at loop edges, and therefore, the effective width of the loop measured in the emission intensity profile is smaller. In such a scenario, one detects effective loop width modulation associated with the periodic re-distribution of the intensity across the loop. \citet{aschwanden2011} reported loop cross-sectional variations in a vertically polarised standing kink mode and interpreted it as a signature of coupled kink and sausage mode. In our simulation, we predict that loop width oscillates at a similar amplitude, but with half the period of the kink mode. According to our modelling, \citet{aschwanden2011} may have observed an overlap of a steady loop and an oscillating loop of similar density and temperature distribution. Therefore, the loop width variation could be accurately measured. The second paper \citep{yuan2016b} in this series will present the modelling details of this event.

In our loop system, the plasma emission of the coronal loop is about two orders of magnitude larger than the background. If the background emission becomes comparable to that of the loop, the spectroscopic measurement is still valid to some extent \citep{yuan2015fm}. However, one may opt to use another spectral line that is much more sensitive to the plasma emission of interest.  

The resolution of the forward models in each view is better than current instruments, i.e., Hinode/EIS. Therefore, to predict the possible observations with EIS, we degrade the resolution to EIS level ($1\arcsec$) by averaging with the neighbouring pixel. The red time series in Figures \ref{fig:top_fe171}, \ref{fig:front_fe171}, \ref{fig:oblique_fe171}, and \ref{fig:side_fe171}, represent the possible sit-and-stare observations with EIS. The loop width is generally measured to be smaller with low-resolution instruments; while the other parameters appear to be a smoothed version of those measured with high-resolution instruments, e.g., the SPICE instrument onboard the Solar Orbiter.

In this study, we only consider a specific case of a standing kink wave in a coronal loop and synthesise the \feix emissions. However, it should be generally applicable to other optically thin emission lines, because in the kink MHD mode, the perturbations to the density and temperature are very tiny. Therefore the spatial distribution of the velocity field plays a determining role in the observational signatures.

\acknowledgements
The research was supported by an Odysseus grant of the FWO Vlaanderen, the IAP P7/08 CHARM (Belspo), the Topping-Up grant CorSeis, the GOA-2015-014 (KU~Leuven), and the Open Research Program KLSA201504 of Key Laboratory of Solar Activity of National Astronomical Observatories of China (D.Y.). CHIANTI is a collaborative project involving George Mason University, the University of Michigan (USA) and the University of Cambridge (UK).

\bibliographystyle{apj}
\bibliography{yuan2015sv}

\begin{thebibliography}{}
\expandafter\ifx\csname natexlab\endcsname\relax\def\natexlab#1{#1}\fi

\bibitem[{{Anfinogentov} {et~al.}(2013){Anfinogentov}, {Nistic{\`o}}, \&
  {Nakariakov}}]{anfinogentov2013}
{Anfinogentov}, S., {Nistic{\`o}}, G., \& {Nakariakov}, V.~M. 2013, \aap, 560,
  A107

\bibitem[{{Antolin} {et~al.}(2015){Antolin}, {Okamoto}, {De Pontieu},
  {Uitenbroek}, {Van Doorsselaere}, \& {Yokoyama}}]{antolin2015}
{Antolin}, P., {Okamoto}, T.~J., {De Pontieu}, B., {et~al.} 2015, \apj, 809, 72

\bibitem[{{Antolin} \& {Van Doorsselaere}(2013)}]{antolin2013}
{Antolin}, P., \& {Van Doorsselaere}, T. 2013, \aap, 555, A74

\bibitem[{{Antolin} \& {Verwichte}(2011)}]{antolin2011}
{Antolin}, P., \& {Verwichte}, E. 2011, \apj, 736, 121

\bibitem[{{Antolin} {et~al.}(2014){Antolin}, {Yokoyama}, \& {Van
  Doorsselaere}}]{antolin2014}
{Antolin}, P., {Yokoyama}, T., \& {Van Doorsselaere}, T. 2014, \apjl, 787, L22

\bibitem[{Arregui(2015)}]{arregui2015}
Arregui, I. 2015, Philosophical Transactions of the Royal Society of London A:
  Mathematical, Physical and Engineering Sciences, 373,
  doi:10.1098/rsta.2014.0261

\bibitem[{{Arregui} {et~al.}(2012){Arregui}, {Oliver}, \&
  {Ballester}}]{arregui2012}
{Arregui}, I., {Oliver}, R., \& {Ballester}, J.~L. 2012, Living Reviews in
  Solar Physics, 9, 2

\bibitem[{{Aschwanden} \& {Boerner}(2011)}]{aschwanden2011b}
{Aschwanden}, M.~J., \& {Boerner}, P. 2011, \apj, 732, 81

\bibitem[{{Aschwanden} {et~al.}(2002){Aschwanden}, {de Pontieu}, {Schrijver},
  \& {Title}}]{aschwanden2002}
{Aschwanden}, M.~J., {de Pontieu}, B., {Schrijver}, C.~J., \& {Title}, A.~M.
  2002, \solphys, 206, 99

\bibitem[{{Aschwanden} {et~al.}(1999){Aschwanden}, {Fletcher}, {Schrijver}, \&
  {Alexander}}]{aschwanden1999}
{Aschwanden}, M.~J., {Fletcher}, L., {Schrijver}, C.~J., \& {Alexander}, D.
  1999, \apj, 520, 880

\bibitem[{{Aschwanden} \& {Nightingale}(2005)}]{aschwanden2005}
{Aschwanden}, M.~J., \& {Nightingale}, R.~W. 2005, \apj, 633, 499

\bibitem[{{Aschwanden} \& {Schrijver}(2011)}]{aschwanden2011}
{Aschwanden}, M.~J., \& {Schrijver}, C.~J. 2011, \apj, 736, 102

\bibitem[{{Brooks} {et~al.}(2013){Brooks}, {Warren}, {Ugarte-Urra}, \&
  {Winebarger}}]{brooks2013}
{Brooks}, D.~H., {Warren}, H.~P., {Ugarte-Urra}, I., \& {Winebarger}, A.~R.
  2013, \apjl, 772, L19

\bibitem[{{Chen} \& {Peter}(2015)}]{chen2015}
{Chen}, F., \& {Peter}, H. 2015, \aap, 581, A137

\bibitem[{{Chen} {et~al.}(2011){Chen}, {Feng}, {Li}, {Song}, {Xia}, {Kong}, \&
  {Li}}]{chen2011}
{Chen}, Y., {Feng}, S.~W., {Li}, B., {et~al.} 2011, \apj, 728, 147

\bibitem[{{Chen} {et~al.}(2010){Chen}, {Song}, {Li}, {Xia}, {Wu}, {Fu}, \&
  {Li}}]{chen2010}
{Chen}, Y., {Song}, H.~Q., {Li}, B., {et~al.} 2010, \apj, 714, 644

\bibitem[{{Cooper} {et~al.}(2003{\natexlab{a}}){Cooper}, {Nakariakov}, \&
  {Tsiklauri}}]{cooper2003a}
{Cooper}, F.~C., {Nakariakov}, V.~M., \& {Tsiklauri}, D. 2003{\natexlab{a}},
  \aap, 397, 765

\bibitem[{{Cooper} {et~al.}(2003{\natexlab{b}}){Cooper}, {Nakariakov}, \&
  {Williams}}]{cooper2003b}
{Cooper}, F.~C., {Nakariakov}, V.~M., \& {Williams}, D.~R. 2003{\natexlab{b}},
  \aap, 409, 325

\bibitem[{{De Moortel} \& {Bradshaw}(2008)}]{demoortel2008}
{De Moortel}, I., \& {Bradshaw}, S.~J. 2008, \solphys, 252, 101

\bibitem[{{De Moortel} {et~al.}(2002{\natexlab{a}}){De Moortel}, {Hood},
  {Ireland}, \& {Walsh}}]{demoortel2002b}
{De Moortel}, I., {Hood}, A.~W., {Ireland}, J., \& {Walsh}, R.~W.
  2002{\natexlab{a}}, \solphys, 209, 89

\bibitem[{{De Moortel} {et~al.}(2002{\natexlab{b}}){De Moortel}, {Ireland},
  {Walsh}, \& {Hood}}]{demoortel2002a}
{De Moortel}, I., {Ireland}, J., {Walsh}, R.~W., \& {Hood}, A.~W.
  2002{\natexlab{b}}, \solphys, 209, 61

\bibitem[{{De Moortel} \& {Nakariakov}(2012)}]{demoortel2012}
{De Moortel}, I., \& {Nakariakov}, V.~M. 2012, Royal Society of London
  Philosophical Transactions Series A, 370, 3193

\bibitem[{{De Moortel} \& {Pascoe}(2009)}]{demoortel2009}
{De Moortel}, I., \& {Pascoe}, D.~J. 2009, \apjl, 699, L72

\bibitem[{{De Pontieu} {et~al.}(2012){De Pontieu}, {Carlsson}, {Rouppe van der
  Voort}, {Rutten}, {Hansteen}, \& {Watanabe}}]{depontieu2012}
{De Pontieu}, B., {Carlsson}, M., {Rouppe van der Voort}, L.~H.~M., {et~al.}
  2012, \apjl, 752, L12

\bibitem[{{Dere} {et~al.}(1997){Dere}, {Landi}, {Mason}, {Monsignori Fossi}, \&
  {Young}}]{dere1997}
{Dere}, K.~P., {Landi}, E., {Mason}, H.~E., {Monsignori Fossi}, B.~C., \&
  {Young}, P.~R. 1997, \aaps, 125, 149

\bibitem[{{DLMS}(2015)}]{NIST:DLMF}
{DLMS}. 2015, {NIST Digital Library of Mathematical Functions},
  http://dlmf.nist.gov/, Release 1.0.9 of 2014-08-29, online companion to
  \cite{Olver:2010:NLMF}

\bibitem[{{Doschek} {et~al.}(2007){Doschek}, {Mariska}, {Warren}, {Brown},
  {Culhane}, {Hara}, {Watanabe}, {Young}, \& {Mason}}]{doschek2007}
{Doschek}, G.~A., {Mariska}, J.~T., {Warren}, H.~P., {et~al.} 2007, \apjl, 667,
  L109

\bibitem[{{Edwin} \& {Roberts}(1982)}]{edwin1982}
{Edwin}, P.~M., \& {Roberts}, B. 1982, \solphys, 76, 239

\bibitem[{{Edwin} \& {Roberts}(1983)}]{edwin1983}
---. 1983, \solphys, 88, 179

\bibitem[{{Erd{\'e}lyi} \& {Morton}(2009)}]{erdelyi2009}
{Erd{\'e}lyi}, R., \& {Morton}, R.~J. 2009, \aap, 494, 295

\bibitem[{{Fang} {et~al.}(2015){Fang}, {Yuan}, {Van Doorsselaere}, {Keppens},
  \& {Xia}}]{fang2015}
{Fang}, X., {Yuan}, D., {Van Doorsselaere}, T., {Keppens}, R., \& {Xia}, C.
  2015, \apj, 813, 33

\bibitem[{{Goossens} {et~al.}(2012){Goossens}, {Andries}, {Soler}, {Van
  Doorsselaere}, {Arregui}, \& {Terradas}}]{goossens2012}
{Goossens}, M., {Andries}, J., {Soler}, R., {et~al.} 2012, \apj, 753, 111

\bibitem[{{Goossens} {et~al.}(2014){Goossens}, {Soler}, {Terradas}, {Van
  Doorsselaere}, \& {Verth}}]{goossens2014}
{Goossens}, M., {Soler}, R., {Terradas}, J., {Van Doorsselaere}, T., \&
  {Verth}, G. 2014, \apj, 788, 9

\bibitem[{{Gruszecki} {et~al.}(2012){Gruszecki}, {Nakariakov}, \& {Van
  Doorsselaere}}]{gruszecki2012}
{Gruszecki}, M., {Nakariakov}, V.~M., \& {Van Doorsselaere}, T. 2012, \aap,
  543, A12

\bibitem[{{Guo} {et~al.}(2015){Guo}, {Ding}, \& {Chen}}]{guo2015}
{Guo}, Y., {Ding}, M.~D., \& {Chen}, P.~F. 2015, \apjs, 219, 36

\bibitem[{{He} {et~al.}(2009){He}, {Marsch}, {Tu}, \& {Tian}}]{he2009}
{He}, J., {Marsch}, E., {Tu}, C., \& {Tian}, H. 2009, \apjl, 705, L217

\bibitem[{{Hershaw} {et~al.}(2011){Hershaw}, {Foullon}, {Nakariakov}, \&
  {Verwichte}}]{hershaw2011}
{Hershaw}, J., {Foullon}, C., {Nakariakov}, V.~M., \& {Verwichte}, E. 2011,
  \aap, 531, A53

\bibitem[{{Heyvaerts} \& {Priest}(1983)}]{heyvaerts1983}
{Heyvaerts}, J., \& {Priest}, E.~R. 1983, \aap, 117, 220

\bibitem[{{Hood} {et~al.}(2009){Hood}, {Browning}, \& {van der
  Linden}}]{hood2009}
{Hood}, A.~W., {Browning}, P.~K., \& {van der Linden}, R.~A.~M. 2009, \aap,
  506, 913

\bibitem[{{Jess} {et~al.}(2015){Jess}, {Morton}, {Verth}, {Fedun}, {Grant}, \&
  {Giagkiozis}}]{jess2015}
{Jess}, D.~B., {Morton}, R.~J., {Verth}, G., {et~al.} 2015, \ssr, 190, 103

\bibitem[{{Kim} {et~al.}(2014){Kim}, {Nakariakov}, \& {Cho}}]{kim2014}
{Kim}, S., {Nakariakov}, V.~M., \& {Cho}, K.-S. 2014, \apjl, 797, L22

\bibitem[{{Klimchuk}(2006)}]{klimchuk2006}
{Klimchuk}, J.~A. 2006, \solphys, 234, 41

\bibitem[{{Kumar} {et~al.}(2013){Kumar}, {Innes}, \& {Inhester}}]{kumar2013}
{Kumar}, P., {Innes}, D.~E., \& {Inhester}, B. 2013, \apjl, 779, L7

\bibitem[{{Kumar} {et~al.}(2015){Kumar}, {Nakariakov}, \& {Cho}}]{kumar2015}
{Kumar}, P., {Nakariakov}, V.~M., \& {Cho}, K.-S. 2015, \apj, 804, 4

\bibitem[{{Kuridze} {et~al.}(2012){Kuridze}, {Morton}, {Erd{\'e}lyi},
  {Dorrian}, {Mathioudakis}, {Jess}, \& {Keenan}}]{kuridze2012}
{Kuridze}, D., {Morton}, R.~J., {Erd{\'e}lyi}, R., {et~al.} 2012, \apj, 750, 51

\bibitem[{{Kuznetsov} {et~al.}(2015){Kuznetsov}, {Van Doorsselaere}, \&
  {Reznikova}}]{kuznetsov2015}
{Kuznetsov}, A.~A., {Van Doorsselaere}, T., \& {Reznikova}, V.~E. 2015,
  \solphys, 290, 1173

\bibitem[{{Lee} {et~al.}(2015){Lee}, {Moon}, \& {Nakariakov}}]{lee2015}
{Lee}, H., {Moon}, Y.-J., \& {Nakariakov}, V.~M. 2015, \apjl, 803, L7

\bibitem[{{Lin} {et~al.}(2007){Lin}, {Engvold}, {Rouppe van der Voort}, \& {van
  Noort}}]{lin2007}
{Lin}, Y., {Engvold}, O., {Rouppe van der Voort}, L.~H.~M., \& {van Noort}, M.
  2007, \solphys, 246, 65

\bibitem[{{Lin} {et~al.}(2009){Lin}, {Soler}, {Engvold}, {Ballester},
  {Langangen}, {Oliver}, \& {Rouppe van der Voort}}]{lin2009}
{Lin}, Y., {Soler}, R., {Engvold}, O., {et~al.} 2009, \apj, 704, 870

\bibitem[{{Liu} {et~al.}(2010){Liu}, {Nitta}, {Schrijver}, {Title}, \&
  {Tarbell}}]{liu2010}
{Liu}, W., {Nitta}, N.~V., {Schrijver}, C.~J., {Title}, A.~M., \& {Tarbell},
  T.~D. 2010, \apjl, 723, L53

\bibitem[{{Liu} \& {Ofman}(2014)}]{liu2014}
{Liu}, W., \& {Ofman}, L. 2014, \solphys, 289, 3233

\bibitem[{{Liu} {et~al.}(2012){Liu}, {Ofman}, {Nitta}, {Aschwanden},
  {Schrijver}, {Title}, \& {Tarbell}}]{liu2012}
{Liu}, W., {Ofman}, L., {Nitta}, N.~V., {et~al.} 2012, \apj, 753, 52

\bibitem[{{Morton}(2014)}]{morton2014}
{Morton}, R.~J. 2014, \aap, 566, A90

\bibitem[{{Morton} \& {McLaughlin}(2013)}]{morton2013}
{Morton}, R.~J., \& {McLaughlin}, J.~A. 2013, \aap, 553, L10

\bibitem[{{Morton} {et~al.}(2015){Morton}, {Tomczyk}, \& {Pinto}}]{morton2015}
{Morton}, R.~J., {Tomczyk}, S., \& {Pinto}, R. 2015, Nature Communications, 6,
  7813

\bibitem[{{Nakariakov} {et~al.}(2009){Nakariakov}, {Aschwanden}, \& {van
  Doorsselaere}}]{nakariakov2009}
{Nakariakov}, V.~M., {Aschwanden}, M.~J., \& {van Doorsselaere}, T. 2009, \aap,
  502, 661

\bibitem[{{Nakariakov} \& {Ofman}(2001)}]{nakariakov2001}
{Nakariakov}, V.~M., \& {Ofman}, L. 2001, \aap, 372, L53

\bibitem[{{Nakariakov} {et~al.}(1999){Nakariakov}, {Ofman}, {Deluca},
  {Roberts}, \& {Davila}}]{nakariakov1999}
{Nakariakov}, V.~M., {Ofman}, L., {Deluca}, E.~E., {Roberts}, B., \& {Davila},
  J.~M. 1999, Science, 285, 862

\bibitem[{{Nakariakov} \& {Verwichte}(2005)}]{nakariakov2005}
{Nakariakov}, V.~M., \& {Verwichte}, E. 2005, Living Reviews in Solar Physics,
  2, 3

\bibitem[{{Nistic{\`o}} {et~al.}(2014{\natexlab{a}}){Nistic{\`o}},
  {Anfinogentov}, \& {Nakariakov}}]{nistico2014b}
{Nistic{\`o}}, G., {Anfinogentov}, S., \& {Nakariakov}, V.~M.
  2014{\natexlab{a}}, \aap, 570, A84

\bibitem[{{Nistic{\`o}} {et~al.}(2013){Nistic{\`o}}, {Nakariakov}, \&
  {Verwichte}}]{nistico2013}
{Nistic{\`o}}, G., {Nakariakov}, V.~M., \& {Verwichte}, E. 2013, \aap, 552, A57

\bibitem[{{Nistic{\`o}} {et~al.}(2014{\natexlab{b}}){Nistic{\`o}}, {Pascoe}, \&
  {Nakariakov}}]{nistico2014}
{Nistic{\`o}}, G., {Pascoe}, D.~J., \& {Nakariakov}, V.~M. 2014{\natexlab{b}},
  \aap, 569, A12

\bibitem[{{Ofman} \& {Thompson}(2002)}]{ofman2002}
{Ofman}, L., \& {Thompson}, B.~J. 2002, \apj, 574, 440

\bibitem[{{Okamoto} {et~al.}(2015){Okamoto}, {Antolin}, {De Pontieu},
  {Uitenbroek}, {Van Doorsselaere}, \& {Yokoyama}}]{okamoto2015}
{Okamoto}, T.~J., {Antolin}, P., {De Pontieu}, B., {et~al.} 2015, \apj, 809, 71

\bibitem[{{Okamoto} \& {De Pontieu}(2011)}]{okamoto2011}
{Okamoto}, T.~J., \& {De Pontieu}, B. 2011, \apjl, 736, L24

\bibitem[{Olver {et~al.}(2010)Olver, Lozier, Boisvert, \&
  Clark}]{Olver:2010:NLMF}
Olver, F.~W.~J., Lozier, D.~W., Boisvert, R.~F., \& Clark, C.~W., eds. 2010,
  {NIST Handbook of Mathematical Functions} (New York, NY: Cambridge University
  Press), print companion to \cite{NIST:DLMF}

\bibitem[{{O'Shea} {et~al.}(2007){O'Shea}, {Srivastava}, {Doyle}, \&
  {Banerjee}}]{oshea2007}
{O'Shea}, E., {Srivastava}, A.~K., {Doyle}, J.~G., \& {Banerjee}, D. 2007,
  \aap, 473, L13

\bibitem[{{Pascoe} {et~al.}(2012){Pascoe}, {Hood}, {de Moortel}, \&
  {Wright}}]{pascoe2012}
{Pascoe}, D.~J., {Hood}, A.~W., {de Moortel}, I., \& {Wright}, A.~N. 2012,
  \aap, 539, A37

\bibitem[{{Pascoe} {et~al.}(2013){Pascoe}, {Nakariakov}, \&
  {Kupriyanova}}]{pascoe2013}
{Pascoe}, D.~J., {Nakariakov}, V.~M., \& {Kupriyanova}, E.~G. 2013, \aap, 560,
  A97

\bibitem[{{Pascoe} {et~al.}(2010){Pascoe}, {Wright}, \& {De
  Moortel}}]{pascoe2010}
{Pascoe}, D.~J., {Wright}, A.~N., \& {De Moortel}, I. 2010, \apj, 711, 990

\bibitem[{{Pascoe} {et~al.}(2011){Pascoe}, {Wright}, \& {De
  Moortel}}]{pascoe2011}
---. 2011, \apj, 731, 73

\bibitem[{{Peter} {et~al.}(2013){Peter}, {Bingert}, {Klimchuk}, {de Forest},
  {Cirtain}, {Golub}, {Winebarger}, {Kobayashi}, \& {Korreck}}]{peter2013}
{Peter}, H., {Bingert}, S., {Klimchuk}, J.~A., {et~al.} 2013, \aap, 556, A104

\bibitem[{{Reale}(2014)}]{reale2014}
{Reale}, F. 2014, Living Reviews in Solar Physics, 11, 4

\bibitem[{{Reznikova} {et~al.}(2014){Reznikova}, {Antolin}, \& {Van
  Doorsselaere}}]{reznikova2014}
{Reznikova}, V.~E., {Antolin}, P., \& {Van Doorsselaere}, T. 2014, \apj, 785,
  86

\bibitem[{{Reznikova} {et~al.}(2015){Reznikova}, {Van Doorsselaere}, \&
  {Kuznetsov}}]{reznikova2015}
{Reznikova}, V.~E., {Van Doorsselaere}, T., \& {Kuznetsov}, A.~A. 2015, \aap,
  575, A47

\bibitem[{{Ruderman}(2003)}]{ruderman2003}
{Ruderman}, M.~S. 2003, \aap, 409, 287

\bibitem[{{Ruderman}(2009)}]{ruderman2009b}
---. 2009, \aap, 506, 885

\bibitem[{{Ruderman} \& {Erd{\'e}lyi}(2009)}]{ruderman2009}
{Ruderman}, M.~S., \& {Erd{\'e}lyi}, R. 2009, \ssr, 149, 199

\bibitem[{{Ruderman} \& {Roberts}(2002)}]{ruderman2002}
{Ruderman}, M.~S., \& {Roberts}, B. 2002, \apj, 577, 475

\bibitem[{{Schrijver} {et~al.}(2002){Schrijver}, {Aschwanden}, \&
  {Title}}]{schrijver2002}
{Schrijver}, C.~J., {Aschwanden}, M.~J., \& {Title}, A.~M. 2002, \solphys, 206,
  69

\bibitem[{{Scullion} {et~al.}(2014){Scullion}, {Rouppe van der Voort},
  {Wedemeyer}, \& {Antolin}}]{scullion2014}
{Scullion}, E., {Rouppe van der Voort}, L., {Wedemeyer}, S., \& {Antolin}, P.
  2014, \apj, 797, 36

\bibitem[{{Selwa} {et~al.}(2010){Selwa}, {Murawski}, {Solanki}, \&
  {Ofman}}]{selwa2010}
{Selwa}, M., {Murawski}, K., {Solanki}, S.~K., \& {Ofman}, L. 2010, \aap, 512,
  A76

\bibitem[{{Selwa} {et~al.}(2007){Selwa}, {Murawski}, {Solanki}, \&
  {Wang}}]{selwa2007}
{Selwa}, M., {Murawski}, K., {Solanki}, S.~K., \& {Wang}, T.~J. 2007, \aap,
  462, 1127

\bibitem[{{Selwa} {et~al.}(2011){Selwa}, {Solanki}, \& {Ofman}}]{selwa2011}
{Selwa}, M., {Solanki}, S.~K., \& {Ofman}, L. 2011, \apj, 728, 87

\bibitem[{{Thurgood} {et~al.}(2014){Thurgood}, {Morton}, \&
  {McLaughlin}}]{thurgood2014}
{Thurgood}, J.~O., {Morton}, R.~J., \& {McLaughlin}, J.~A. 2014, \apjl, 790, L2

\bibitem[{{Tomczyk} {et~al.}(2007){Tomczyk}, {McIntosh}, {Keil}, {Judge},
  {Schad}, {Seeley}, \& {Edmondson}}]{tomczyk2007}
{Tomczyk}, S., {McIntosh}, S.~W., {Keil}, S.~L., {et~al.} 2007, Science, 317,
  1192

\bibitem[{{Tripathi} {et~al.}(2009){Tripathi}, {Isobe}, \&
  {Jain}}]{tripathi2009}
{Tripathi}, D., {Isobe}, H., \& {Jain}, R. 2009, \ssr, 149, 283

\bibitem[{{Van Doorsselaere} {et~al.}(2004){Van Doorsselaere}, {Andries},
  {Poedts}, \& {Goossens}}]{vandoorsselaere2004}
{Van Doorsselaere}, T., {Andries}, J., {Poedts}, S., \& {Goossens}, M. 2004,
  \apj, 606, 1223

\bibitem[{{Van Doorsselaere} {et~al.}(2014){Van Doorsselaere}, {Gijsen},
  {Andries}, \& {Verth}}]{vandoorsselaere2014}
{Van Doorsselaere}, T., {Gijsen}, S.~E., {Andries}, J., \& {Verth}, G. 2014,
  \apj, 795, 18

\bibitem[{{Van Doorsselaere} \& {Nakariakov}(2008)}]{vandoorsselaere2008b}
{Van Doorsselaere}, T., \& {Nakariakov}, V.~M. 2008, in Astronomical Society of
  the Pacific Conference Series, Vol. 397, First Results From Hinode, ed. S.~A.
  {Matthews}, J.~M. {Davis}, \& L.~K. {Harra}, 58

\bibitem[{{Van Doorsselaere} {et~al.}(2008){Van Doorsselaere}, {Nakariakov}, \&
  {Verwichte}}]{vandoorsselaere2008}
{Van Doorsselaere}, T., {Nakariakov}, V.~M., \& {Verwichte}, E. 2008, \apjl,
  676, L73

\bibitem[{{Van Doorsselaere} {et~al.}(2009){Van Doorsselaere}, {Verwichte}, \&
  {Terradas}}]{vandoorsselaere2009}
{Van Doorsselaere}, T., {Verwichte}, E., \& {Terradas}, J. 2009, \ssr, 149, 299

\bibitem[{{Verwichte} {et~al.}(2009){Verwichte}, {Aschwanden}, {Van
  Doorsselaere}, {Foullon}, \& {Nakariakov}}]{verwichte2009}
{Verwichte}, E., {Aschwanden}, M.~J., {Van Doorsselaere}, T., {Foullon}, C., \&
  {Nakariakov}, V.~M. 2009, \apj, 698, 397

\bibitem[{{Verwichte} {et~al.}(2006){Verwichte}, {Foullon}, \&
  {Nakariakov}}]{verwichte2006}
{Verwichte}, E., {Foullon}, C., \& {Nakariakov}, V.~M. 2006, \aap, 452, 615

\bibitem[{{Verwichte} {et~al.}(2010){Verwichte}, {Foullon}, \& {Van
  Doorsselaere}}]{verwichte2010}
{Verwichte}, E., {Foullon}, C., \& {Van Doorsselaere}, T. 2010, \apj, 717, 458

\bibitem[{{Verwichte} {et~al.}(2005){Verwichte}, {Nakariakov}, \&
  {Cooper}}]{verwichte2005}
{Verwichte}, E., {Nakariakov}, V.~M., \& {Cooper}, F.~C. 2005, \aap, 430, L65

\bibitem[{{Verwichte} {et~al.}(2004){Verwichte}, {Nakariakov}, {Ofman}, \&
  {Deluca}}]{verwichte2004}
{Verwichte}, E., {Nakariakov}, V.~M., {Ofman}, L., \& {Deluca}, E.~E. 2004,
  \solphys, 223, 77

\bibitem[{{Verwichte} {et~al.}(2013{\natexlab{a}}){Verwichte}, {Van
  Doorsselaere}, {Foullon}, \& {White}}]{verwichte2013}
{Verwichte}, E., {Van Doorsselaere}, T., {Foullon}, C., \& {White}, R.~S.
  2013{\natexlab{a}}, \apj, 767, 16

\bibitem[{{Verwichte} {et~al.}(2013{\natexlab{b}}){Verwichte}, {Van
  Doorsselaere}, {White}, \& {Antolin}}]{verwichte2013b}
{Verwichte}, E., {Van Doorsselaere}, T., {White}, R.~S., \& {Antolin}, P.
  2013{\natexlab{b}}, \aap, 552, A138

\bibitem[{{Wang} {et~al.}(2009{\natexlab{a}}){Wang}, {Ofman}, \&
  {Davila}}]{wang2009a}
{Wang}, T.~J., {Ofman}, L., \& {Davila}, J.~M. 2009{\natexlab{a}}, \apj, 696,
  1448

\bibitem[{{Wang} {et~al.}(2009{\natexlab{b}}){Wang}, {Ofman}, {Davila}, \&
  {Mariska}}]{wang2009b}
{Wang}, T.~J., {Ofman}, L., {Davila}, J.~M., \& {Mariska}, J.~T.
  2009{\natexlab{b}}, \aap, 503, L25

\bibitem[{{Wang} \& {Solanki}(2004)}]{wang2004}
{Wang}, T.~J., \& {Solanki}, S.~K. 2004, \aap, 421, L33

\bibitem[{{White} \& {Verwichte}(2012)}]{white2012}
{White}, R.~S., \& {Verwichte}, E. 2012, \aap, 537, A49

\bibitem[{{White} {et~al.}(2012){White}, {Verwichte}, \&
  {Foullon}}]{white2012b}
{White}, R.~S., {Verwichte}, E., \& {Foullon}, C. 2012, \aap, 545, A129

\bibitem[{{Winebarger} {et~al.}(2002){Winebarger}, {Warren}, {van
  Ballegooijen}, {DeLuca}, \& {Golub}}]{winebarger2002}
{Winebarger}, A.~R., {Warren}, H., {van Ballegooijen}, A., {DeLuca}, E.~E., \&
  {Golub}, L. 2002, \apjl, 567, L89

\bibitem[{{Yuan} \& {Nakariakov}(2012)}]{yuan2012sm}
{Yuan}, D., \& {Nakariakov}, V.~M. 2012, \aap, 543, A9

\bibitem[{{Yuan} {et~al.}(2015{\natexlab{a}}){Yuan}, {Pascoe}, {Nakariakov},
  {Li}, \& {Keppens}}]{yuan2015rs}
{Yuan}, D., {Pascoe}, D.~J., {Nakariakov}, V.~M., {Li}, B., \& {Keppens}, R.
  2015{\natexlab{a}}, \apj, 799, 221

\bibitem[{{Yuan} {et~al.}(2013){Yuan}, {Shen}, {Liu}, {Nakariakov}, {Tan}, \&
  {Huang}}]{yuan2013fw}
{Yuan}, D., {Shen}, Y., {Liu}, Y., {et~al.} 2013, \aap, 554, A144

\bibitem[{{Yuan} \& {Van Doorsselaere}(2016)}]{yuan2016b}
{Yuan}, D., \& {Van Doorsselaere}, T. 2016, ArXiv e-prints, arXiv:1602.07598

\bibitem[{{Yuan} {et~al.}(2015{\natexlab{b}}){Yuan}, {Van Doorsselaere},
  {Banerjee}, \& {Antolin}}]{yuan2015fm}
{Yuan}, D., {Van Doorsselaere}, T., {Banerjee}, D., \& {Antolin}, P.
  2015{\natexlab{b}}, \apj, 807, 98

\bibitem[{{Zimovets} \& {Nakariakov}(2015)}]{zimovets2015}
{Zimovets}, I.~V., \& {Nakariakov}, V.~M. 2015, \aap, 577, A4

\end{thebibliography}
\appendix
\section{Derivation of the quadrupole terms}
\label{sec:appendix}
Here we demonstrate the derivation of \eqref{eq:vx} and \ref{eq:vy}. 
\begin{align}
 \tilde{v}_x&=\hat{v}_r\cos^2\phi-\hat{v}_\phi \sin^2\phi, \\
      &=\frac{-A \omega \kappa_r}{\rho_0(\omega^2-\omega_{\A}^2)}\left(\frac{\d \RR}{\kappa_r \d r}\cos^2\phi+\frac{\RR}{\kappa_r r} \sin^2\phi \right).
\end{align}
We define $\lambda=\kappa_r r$ and $\RR'(\lambda)=\d \RR/\d \lambda$. 
For the plasma motion inside the loop $r<a$: 
\begin{align}
\frac{\d \RR}{\kappa_r \d r}\cos\phi^2+\frac{\RR}{\kappa_r r} \sin\phi^2&= J_1'(\lambda)\cos^2\phi+\frac{J_1(\lambda)}{\lambda}\sin^2\phi \\
 &=(\frac{J_1}{\lambda}-J_2)\cos^2\phi+\frac{J_1(\lambda)}{\lambda}\sin^2\phi \\
 &=\frac{J_1}{\lambda}-J_2\cos^2\phi \\
 &=\frac{J_1}{\lambda}-\frac{J_2}{2}-\frac{J_2}{2}\cos2\phi \\
 &=\frac{J_0-J_2\cos2\phi}{2},
\end{align}
where we used $J_1'(\lambda)=J_1/\lambda-J_2$ and $J_2=2J_1/\lambda-J_0$ \citep{Olver:2010:NLMF} in the derivation.
We follow the same procedure and used $K_1'=K_1/\lambda-K_2$ and $K_2=2K_1/\lambda+K_0$ \citep{Olver:2010:NLMF}. Then we obtained the plasma motion outside the loop $r>a$:
\begin{align}
\frac{\d \RR}{\kappa_r \d r}\cos\phi^2+\frac{\RR}{\kappa_r r} \sin\phi^2&= K_1'(\lambda)\cos^2\phi+\frac{K_1(\lambda)}{\lambda}\sin^2\phi \\
 &=\frac{K_1}{\lambda}-K_2\cos^2\phi \\
 &=-\frac{K_0+K_2\cos2\phi}{2}
\end{align}
Here, we obtain the horizontal motion,
\begin{align}
\tilde{v}_x&=\left\{
	  \begin{array}{lr}
	      \frac{-A_\i \omega \kri}{2\rho_\i(\omega^2-\omega_{\A\i}^2)}(J_0-J_2\cos2\phi),\;\text{for $r \leq a$} \\
	      \frac{-A_\e \omega \kre}{2\rho_\e(\omega^2-\omega_{\A\e}^2)}(-K_0-K_2\cos 2\phi), \;\text{for $r>a$}.
	  \end{array}\right.\\
\end{align}
At $r=0$, $\tilde{v}_x=v_{00}=\frac{-A_\i \omega \kri}{2\rho_\i(\omega^2-\omega_{\A\i}^2)}$; while at $r=a$ and $\phi=0$ or $\pi$,
\begin{align}
 \tilde{v}_x|_{r=a^-}&=v_{00}(J_0(\kri a)-J_2(\kri a)),\; \\
\tilde{v}_x|_{r=a^+} &=\frac{-A_\e \omega \kre}{2\rho_\e(\omega^2-\omega_{\A\e}^2)}(-K_0(\kre a) -K_2 (\kre a)),\\
 &=\frac{J'_1(\kri a)}{K'_1(\kre a)}v_{00}(-K_0(\kre a) -K_2 (\kre a)).
\end{align}
where we recalled the total pressure balance at $r=a$ and the dispersion relationship (\eqref{eq:disp}):
\begin{align}
 A_\i J_1(\kri a) &=  A_\e K_1(\kre a) \\
\frac{\rho_\i(\omega_{\A\i}^2-\omega^2) \kre}{\rho_\e(\omega_{\A\e}^2-\omega^2)\kri}&=\frac{J'_1(\kri a) K_1(\kre a)}{J_1(\kri a) K'_1(\kre a)}.
\end{align}
If we use the relationship $J'_1=0.5(J_0-J_2)$ and $K'_1=-0.5(K_0+K_2)$ \citep{Olver:2010:NLMF}, then we could verify that $\tilde{v}_x|_{r=a^-}=\tilde{v}_x|_{r=a^+}$ at $\phi=0$ and $\pi$.

If one follows the same procedure and uses the relationships \citep[$J'_1=J_1/\eta-J_2$ and $K'_1=K_1/\eta-K_2$,][]{Olver:2010:NLMF}, then one could easily obtain the vertical component:
\begin{align}
 \tilde{v}_y&=\left\{
 	  \begin{array}{lr}
 	      -v_{00}J_2\sin2\phi,\;\text{for $r \leq a$} \\
 	      -\frac{J'_1(|\kappa_{r\i}|a)}{K'_1(\kappa_{r\e}a)}v_{00}K_2\sin 2\phi,\;\text{for $r>a$}.
 	  \end{array}\right.
\end{align}

\end{document}